\documentclass{aastex}
\usepackage{emulateapj5, psfig, epsfig}

\begin{document}

\received{}
\revised{}
\accepted{}

\lefthead{M. Catelan, F.~R. Ferraro, \& R.~T.~Rood}
\righthead{}

\slugcomment{The Astrophysical Journal, in press}

\singlespace

\title{Horizontal-Branch Models and the Second-Parameter Effect.\\
       IV. The Case of M3 and Palomar~3}

\author{M.~Catelan\altaffilmark{1,}\altaffilmark{2}}

\author{F.~R.~Ferraro\altaffilmark{3}}

\and

\author{R.~T.~Rood\altaffilmark{2}}

\altaffiltext{1}{Hubble Fellow.}
\altaffiltext{2}
      {
       University of Virginia,
       Department of Astronomy,
       P.O.~Box~3818,
       Charlottesville, VA 22901-2436, USA;
       e-mail: catelan/rtr@virginia.edu
       }
\altaffiltext{3}
      {
       Osservatorio Astronomico di Bologna,
       via Ranzani~1,
       I-40126 Bologna, Italy;
       e-mail: ferraro@astbo3.bo.astro.it
      }
\begin{abstract}
We present a detailed analysis of the ``second-parameter pair" of 
globular clusters M3 (NGC~5272) and Palomar~3. Our main results can 
be summarized as follows: 
i)~The
horizontal-branch (HB) morphology of M3 is significantly bluer in its
inner regions (observed with the {\em Hubble Space Telescope}) than in
the cluster outskirts (observed from the ground), i.e., M3 has an
internal second parameter. Most plausibly the mass loss on the red
giant branch (RGB) has been more efficient in the inner than in the
outer regions of the cluster. 
ii)~The dispersion in mass of the Pal~3 HB
is found to be very small---consistent with zero---and we argue that 
this is unlikely to be
due to a statistical fluctuation. It is this small mass dispersion
that leads to the most apparent difference in the HB morphologies of
M3 and Pal~3.  
iii)~The relative HB types of M3 and Pal~3, as measured by mean
colors or parameters involving the number of blue, variable, and red
HB stars, can easily be accounted for by a fairly small difference in
age between these clusters, of order 0.5--1~Gyr---which is in good 
agreement with the relative age measurement, based on the clusters'
turnoffs, by VandenBerg (2000). 
\end{abstract}


\section{Introduction}

Many recent studies of the globular clusters (GCs) in the outer halo
of our Galaxy have focused on the age difference between these distant
clusters and those located in the inner halo. Since Mironov \& Samus
(1974) and Searle \& Zinn (1978), it has been widely recognized that
the possible existence of such age differences would have important
implications for our understanding of the early stages of the
formation of the Galaxy (e.g., van den Bergh 1993; Zinn 1993; Lee, 
Demarque, \& Zinn 1994; Majewski 1994). While in these early studies 
the morphology
of the horizontal branch (HB) was assumed to be a safe relative age
indicator---a critical working hypothesis in those days, given the
tremendous challenge that accurate and precise photometry reaching
below distant clusters' turnoffs (the primary stellar evolution
``clock") represented---more recently {\em Hubble Space Telescope}
(HST) photometry of the most distant GCs in the Galaxy have allowed a
direct and precise estimate of the position of their main-sequence
turnoffs, thus enabling the {\em direct} determination of their ages
as compared to closer GCs with good ground-based photometry (Harris
et al. 1997; Stetson et al. 1999; VandenBerg 2000).

As a result, it was found that the most massive of the extreme
outer-halo GCs, NGC~2419, has the same age as the well-studied metal-poor,
inner-halo GC M92~=~NGC~6341 (Harris et al. 1997). It is worth
noting that M92 is often considered to be one of the oldest GCs in the
Galaxy (e.g., Bolte \& Hogan 1995). This result gives support to the
growing notion that the first episodes of star formation occurred
more or less simultaneously throughout most of the Local Group
(van den Bergh 1999, 2000; Grebel 2001; and references therein).

\vskip 0.2in
\begin{figure*}[ht]
 \centerline{\epsfig{file=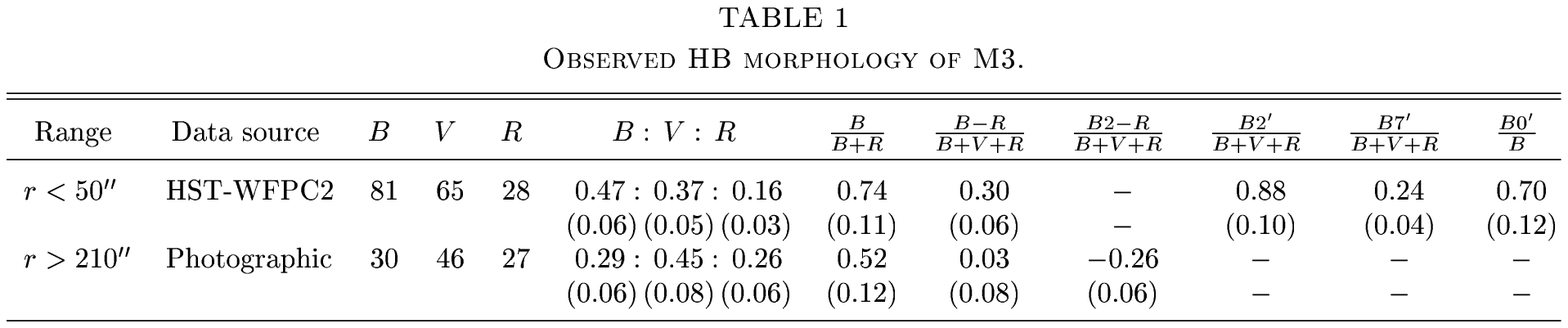,height=1.5in,width=7in}} 
\end{figure*}
\vskip 0.2in

%
\begin{figure*}[t]
 \figurenum{1}
 \caption{
          De-reddened CMDs for the internal (panel a) and external 
          (panel b) regions of M3. The region occupied by RR Lyrae 
          variables is schematically indicated by boxes. The number 
          of blue ($B$), variable ($V$), and red ($R$) HB stars is 
          also indicated; note that, for the same number of red HB 
          stars, one finds 2.7 times as many blue HB stars in the 
          core than in the cluster outskirts. This sharp difference 
          in HB morphology between different radial regions of M3 
          is real, statistically significant and cannot be caused by 
          any spurious problems with the data or the photometry. 
          Two randomly picked 
          CMD simulations for the inner cluster region
          are shown in the $M_V$, $(V-I)_0$ plane in panels
          c) and e). Likewise, two randomly picked CMD simulations 
          for the outer cluster region are shown 
          in the $M_V$, $(\bv)_0$ plane in panels d) and f). 
          Note that the
          total number of stars in the simulations was allowed to
          randomly fluctuate around the values $B+V+R = 174$ and 
          103 (panels c/e and d/f, respectively), following
          the Poisson distribution. Note also that the AGB phase 
          is not included in these models. 
         }
\end{figure*}

On the other hand, recent HST studies of the sparse 
outer-halo GCs Palomar~3, Palomar~4, and Eridanus have indicated
that they, unlike NGC~2419, might be younger than inner-halo GCs
of similar metallicity.\footnote{For the 
definition of ``outer" and ``inner" halo, see the discussion in
\S7 of Borissova et al. (1997).} Since these globulars are also
commonly considered typical cases of loose GCs with predominantly
red HBs, they are often quoted as ``second-parameter" GCs. For instance,
Gratton \& Ortolani (1984) note that the HBs of Pal~3 and Pal~4 ``are
too red for [their] metal abundances," thus confirming ``the need of
a second parameter to explain the morphology of the red HBs of outer
halo clusters" (see also Stetson et al. 1999 and VandenBerg 2000, who 
explicitly compare Pal~3 against M3, on the one hand, and Pal~4/Eridanus 
against M5, on the other hand). 

As widely discussed in the literature, 
perhaps the primary second-parameter candidate is age, though it 
remains to be proved that age is the only, or even the most important, 
second parameter (see, e.g., Stetson, VandenBerg, \& Bolte 1996; 
Sarajedini, Chaboyer, \& Demarque 1997; Fusi Pecci \& Bellazzini 1997; 
Catelan 2000 and references therein). 

This notwithstanding, that age is {\em the} second parameter has been 
a frequently adopted assumption in the astronomical literature, both 
for Galactic and extragalactic GCs (e.g., Sarajedini et al. 1998). 
{\em Therefore, it is of paramount importance to investigate whether 
the observed HB morphology differences are consistent with such a 
hypothesis for systems with detected turnoff age differences.} 
Arguably, this is also a necessary step before we can be in a position 
to reliably explore the role (if any) of other second parameter 
candidates besides age; after all, if age differences can be shown  
to be compatible with the observed differences in HB morphology for 
second-parameter pairs, one might invoke Occam's Razor in support  
of the view that, since no other second parameters besides age are 
{\em required}, then no other second parameters besides age actually 
{\em exist}.

Original work by Ortolani \& Gratton (1989) had previously hinted
that the ``second parameter," in the case of Pal~3 and M3, might
indeed be age. On the other hand, the preliminary HST report by
Richer et al. (1996; see their Table~1) supported an age difference
of $0.0$~Gyr between Pal~3 and M3, which---if confirmed---would
obviously rule out age as the second parameter for this pair.

More recently, both Stetson et al. (1999) and VandenBerg (2000)
conducted much more detailed comparisons of the HST turnoff ages of the
outer-halo GCs Pal~3, Pal~4, and Eridanus with those of M3~=~NGC~5272
(in the case of Pal~3) and M5~=~NGC~5904 (in the cases of Pal~4
and Eridanus). Curiously, the age differences favored by Stetson
et al. and by VandenBerg differ by $\sim 50\%$, even though
they are based on the same HST data---with VandenBerg's analysis
generally supporting smaller age differences for the same
second-parameter pairs. (The reasons for such differences are discussed
in great detail by VandenBerg.) Two additional extreme outer-halo GCs,
Pal~14 and AM-1, were observed by the same team during HST's Cycle~6
(GO-6512, PI~Hesser), but, to the best of our knowledge, no results
of this investigation have been reported in the literature as yet.

In the previous paper of this series (Catelan 2000), we
analyzed the outer-halo GCs Pal~4 and Eridanus in the context
of the ``second-parameter" phenomenon. In particular, we have
quantitatively investigated whether the turnoff ages of these
outer-halo GCs, as compared to M5, are consistent with their HB
morphologies (M5 having by far the bluer HB). The main conclusion
of that study was that the HST turnoff age difference, whether as
measured by Stetson et al. (1999) or by VandenBerg (2000), is too
small to fully explain the second-parameter effect seen in the case
of Pal~4/Eridanus vs. M5.

The purpose of the present paper is to extend Catelan's (2000)
analysis of the outer-halo GCs by carrying out a detailed
investigation of M3 and Pal~3, which we shall assume to be a 
bona-fide ``second-parameter pair"
(Burbidge \& Sandage 1958; Gratton \& Ortolani 1984; Ortolani \&
Gratton 1989; Stetson et al. 1999; VandenBerg 2000). In particular, 
our goal is to answer the following question: Assuming that M3 and 
Pal~3 is a bona-fide second-parameter pair, are the HB morphology 
differences between these clusters consistent with the hypothesis 
that age is {\em the} second parameter of HB morphology?

%
\begin{figure*}[ht]
 \figurenum{2}
 \centerline{\psfig{figure=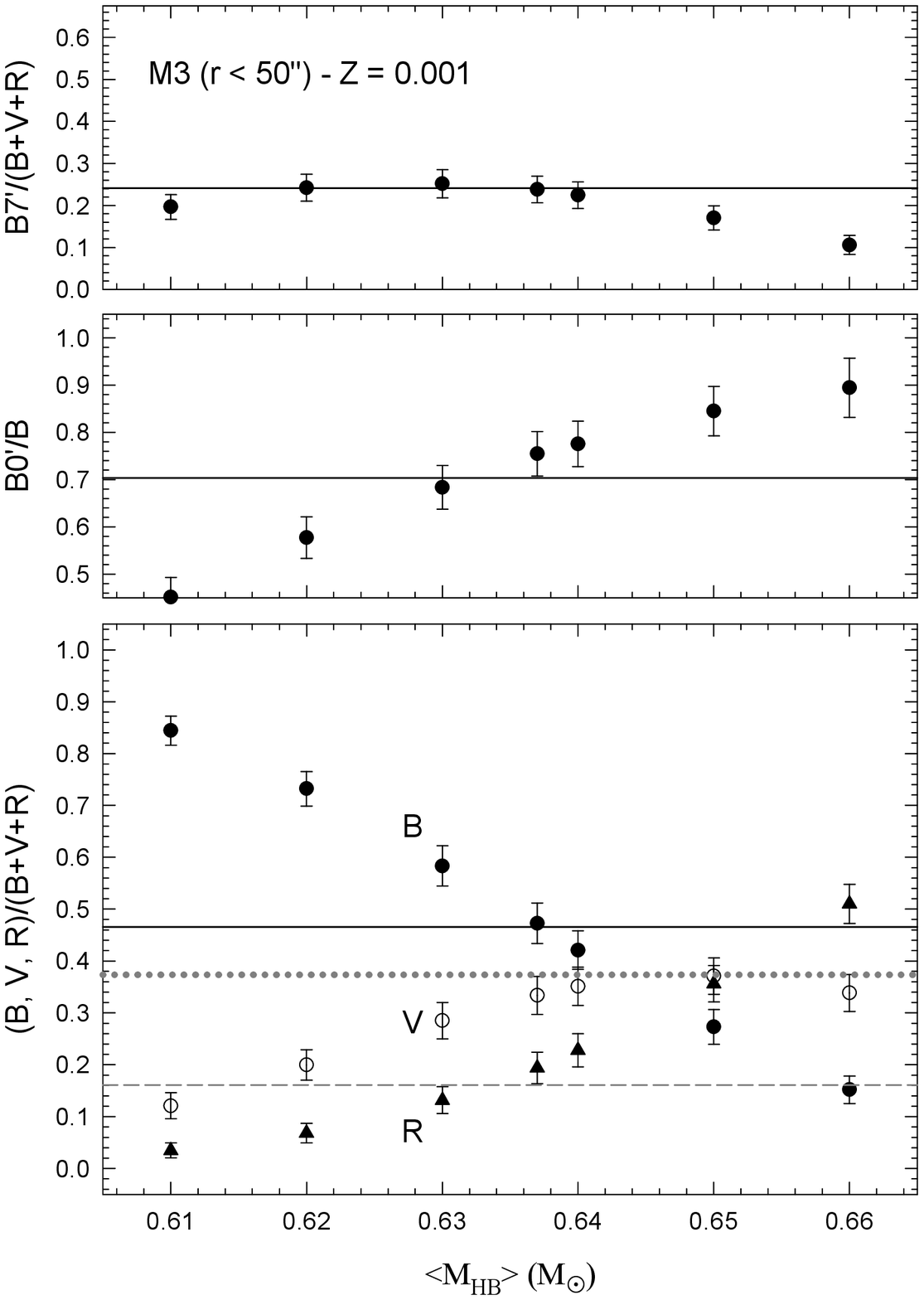}}
 \caption{Illustration of the dependence of several HB morphology 
          parameters on $\langle M_{\rm HB}\rangle$ for M3. 
          The mass dispersion parameter was held fixed at a value 
          $\sigma_M = 0.023 \, M_{\odot}$, appropriate for the  
          {\em inner cluster regions} ($r < 50\arcsec$).  
          Each datapoint was obtained from an average of 1000 HB 
          simulations. Horizontal lines indicate the observed 
          values for M3. In the bottom panel, the solid line 
          indicates $B/(B+V+R)$, whereas dotted and dashed gray 
          lines denote $V/(B+V+R)$ and $R/(B+V+R)$, respectively. 
          In the the top panel, $B7\arcmin/(B+V+R)$ is given, 
          while the middle panel shows $B0\arcmin/B$. 
         }
\end{figure*}

%
\begin{figure*}[t]
 \figurenum{3}
 \centerline{\psfig{figure=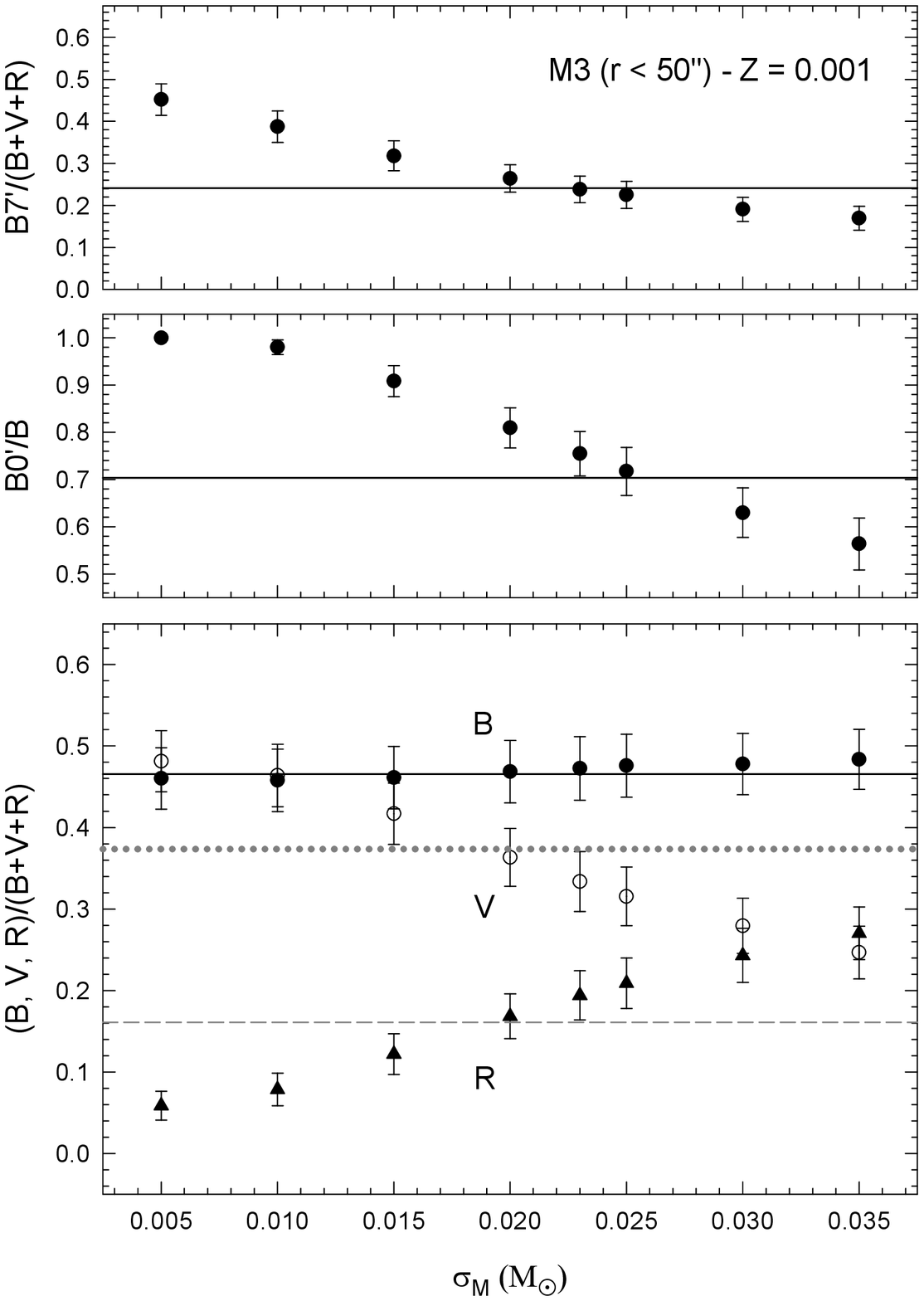}}
 \caption{As in Fig.~2, but for $\sigma_M$. In these computations,  
          the mean HB mass was held fixed at a value 
          $\langle M_{\rm HB}\rangle = 0.637 \, M_{\odot}$, as 
          found appropriate for M3's {\em inner regions} 
          ($r < 50\arcsec$). 
         }
\end{figure*}

%
\begin{figure*}[t]
 \figurenum{4}
 \centerline{\psfig{figure=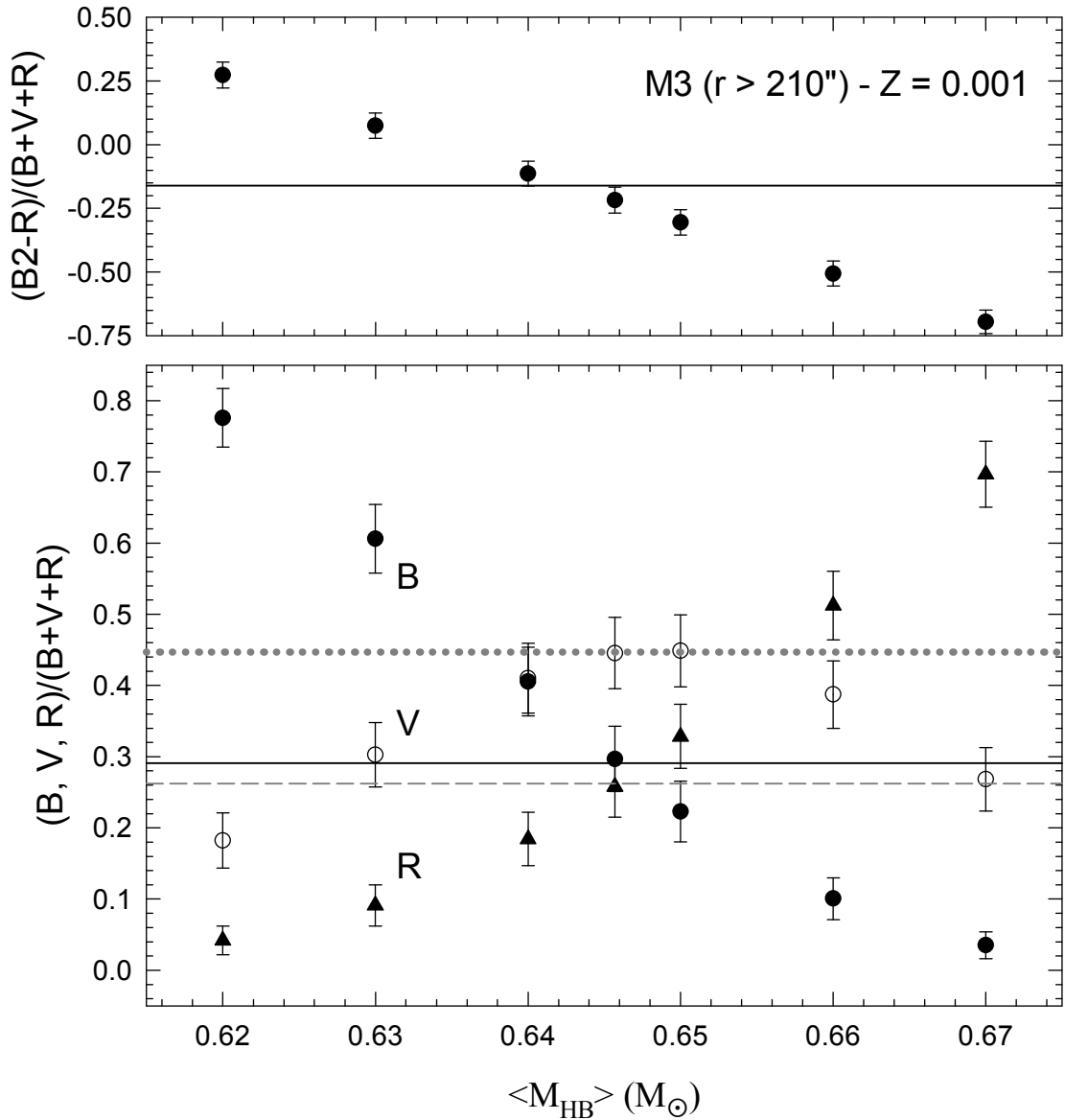}}
 \caption{Similar to Fig.~2, but for M3's {\em outer regions} 
          ($r > 210\arcsec$); in these computations, 
          the mass dispersion parameter was accordingly held fixed  
          at a value $\sigma_M = 0.018 \, M_{\odot}$. 
          The upper panel shows the behavior of the Buonanno (1993)  
          parameter, $(B2-R)/(B+V+R)$. 
         }
\end{figure*}

%
\begin{figure*}[ht]
 \figurenum{5}
 \centerline{\psfig{figure=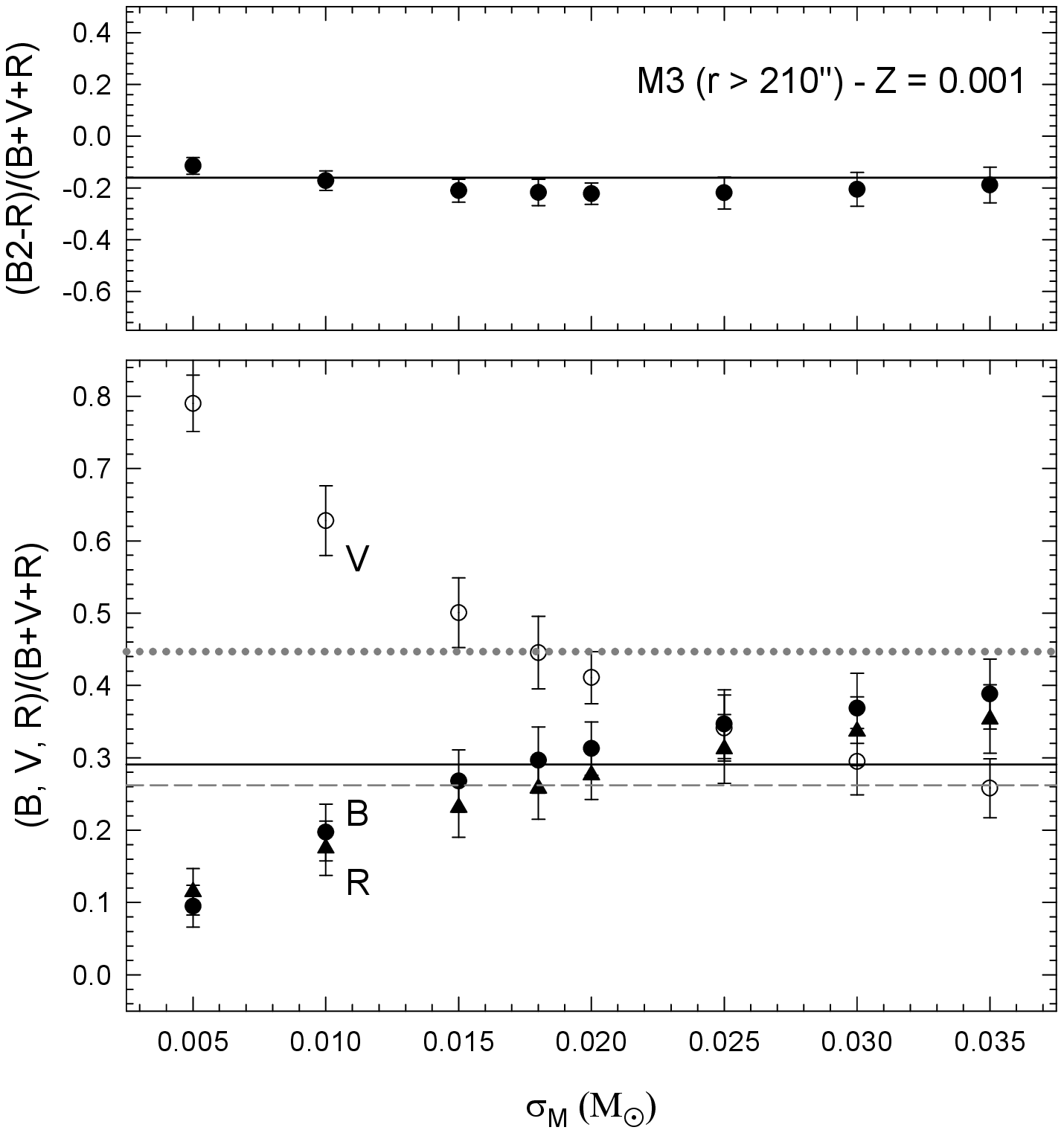}}
 \caption{As in Fig.~4, but for $\sigma_M$. In these computations,  
          the mean HB mass was held fixed at a value 
          $\langle M_{\rm HB}\rangle = 0.646 \, M_{\odot}$, as 
          found appropriate for M3's {\em outer regions} 
          ($r > 210\arcsec$). 
         }
\end{figure*}

We begin in the next section by describing the HB morphology of M3
as a function of distance from the cluster center, showing that the
HB of the cluster is significantly bluer in its center than in the
outermost regions. The CMD of Pal~3, as obtained from the HST-WFPC2
data of Stetson et al. (1999), is also presented and discussed;
previous ground-based photometry of the cluster is also briefly
addressed. In \S3, we build synthetic models of the HBs of the two
clusters, and present our best-fitting solutions. In \S4, the relative
turnoff ages of M3 and Pal~3, as determined by Stetson et al. and by
VandenBerg (2000), are compared with the relative ages that are
required to account for their different HB
types, under the assumption that age is the second parameter of HB
morphology. Our concluding remarks, with emphasis on the evidence
of ``local" second parameters leading to the most apparent difference 
in HB morphology between M3 and Pal~3, are finally presented in \S5.

\section{Observational Data}

\subsection{HB Morphology of M3}

During the last 10 years we collected and published a large
photometric data base on M3. Using different techniques
(photographic plates: Buonanno et al. 1994; ground-based
CCD photometry: Ferraro et al. 1993, Ferraro et al. 1997a,
Carretta et al. 1998; HST-WFPC2: Ferraro et al. 1997b),
we secured a huge sample covering the entire cluster extension
from the very center of the cluster up to $\sim 7\arcmin$.
On the basis of this dataset we constructed one of the largest
and most complete luminosity functions for evolved stars
(brighter than the main-sequence turnoff) ever published for
a GC (Rood et al. 1999).

Using such a dataset we investigated the HB morphology in M3
and
discovered that it turns out to be significantly bluer
in the inner region than in the outskirts.
While a full paper (in preparation) is devoted to accurately 
discussing the HB morphology variation as a function of the distance 
from the cluster center, here we want to point out the sharp difference 
in the HB star distribution between the innermost and the outermost 
regions.
Figure~1 shows the M3 CMDs for the innermost 
($r<50\arcsec$, panel a) and the outermost ($r>210\arcsec$, panel b) 
samples.
A box has been plotted to schematically indicate the location of the 
variable (RR Lyrae--type) stars; the corresponding 
$V$ number is also provided. 
Figs.~1c--f, as well as Figs.~2--5, 
refer to our theoretical modelling of the cluster, and will be 
discussed in \S3.1.

More information on population ratios is 
summarized in Table~1. In column~1, the radial range is given. In
column~2, the type of data utilized (HST-WFPC2, ground-based  
photographic photometry) is indicated.
In columns~3, 4, and 5, one finds $B$, $V$, and $R$ values,
respectively. The corresponding
$B:\,V:\,R$ ratios are listed in column~6. In columns 7 through 9, 
the values of the Mironov (1972), Lee--Zinn (Zinn 1986; Lee et al. 
1994), and Buonanno (1993) HB morphology parameters, respectively, 
are given. The last three columns provide the values of some  
additional HB morphology parameters defined by Catelan et al. (2001); 
we briefly recall that $B2\arcmin$ represents the 
number of HB stars redder than $(V-I)_0 = -0.02$, whereas
$B7\arcmin$ and $B0\arcmin$ represent the number of {\em blue} HB
stars redder than $(V-I)_0 = +0.07$ and $0$, respectively.  

As can be seen from Fig.~1, in selecting 
different groups along the HB we followed the criteria
described in Buonanno et al. (1994) and Ferraro et al. (1997a):
in particular, both ``Extreme Blue" (EB), with $V>17$, and ``Extreme
Red" (ER), with $(\bv)_0 > 0.6$ or $(V-I)_0 > 0.8$, HB stars have been
excluded from the $B$ and $R$ number counts, respectively. In fact,
these stars might be ``non genuine" HB stars, in the sense that 
they may derive from different (non-canonical) 
phenomena: ER objects might represent the progeny of the large
blue straggler population discovered in this cluster (Ferraro et al.
1993, 1997b); and EB stars could derive from (say) late collisonal 
mergers (Bailyn \& Iben 1989).
In any event, ER and EB stars represent a very minor fraction of the
total HB population of M3---3.3\% and 2.6\%, respectively---so that
the results of this paper are not seriously affected by our not
having included them in the number counts reported in Table~1.

As can be seen from the data plotted in Figs.~1a,b and summarized 
in Table~1, the detected difference in the M3 HB morphology is
sharp and significant. It is difficult to attribute it to any 
spurious effects (due to crowding, etc.) or to any additional 
source of error associated with the photometry. Furthermore, the 
limiting magnitude of the M3 photometry is several magnitudes below 
the instability strip level in Figs.~1a,b, ruling out significant 
differential incompleteness in HB star detection from one side of 
the HB to the other. 
In our view this represents clear-cut evidence that a mechanism 
able to modify the
star distribution along the HB is at work in M3 and  
is more effective in the central region of the cluster.
The problem  is surely worthy of further investigation since 
anomalous radial behaviour has been seen before in other stellar 
populations in this cluster (see the case of the blue straggler radial 
distribution discussed in Ferraro et al. 1993 and Ferraro et al. 1997b). 

\vskip 0.15in
\subsection{HB Morphology of Palomar~3}
The dereddened $V$, $(V-I)_0$
Pal~3 CMD, based on HST-WFPC2 data kindly provided
to us by Dr. P. B. Stetson, is shown in Figs.~6a,b. RR Lyrae variables
are indicated by circled dots. The adopted reddening comes from
the Schlegel, Finkbeiner, \& Davis (1998) dust maps, which give
$E(\bv) = 0.042$~mag at the position of Pal~3, and the assumption
that $E(V-I)/E(\bv)\simeq 1.3$ (see Stetson et al. 1999 and
references therein). The adopted reddening is in excellent
agreement with the value tabulated by Harris (1996) in the 
June~1999 update of his catalog, $E(\bv) = 0.04$~mag---an estimate 
which is completely independent of the Schlegel et al. maps. 
For reference purposes, a vertical dotted line representing
the mean dereddened color of the combined HBs of Pal~4 and Eridanus
[$(V-I)_0 \simeq 0.78$~mag; see Catelan 2000] is overplotted
on the Pal~3 CMD of Figs.~6a,b. In Fig.~6b, the zero-age HB for 
$Z = 0.001$ is overplotted on the data, as is an evolutionary 
track for a $0.656\,M_{\odot}$ star, after appropriately applying 
a shift in $V$ by 19.67~mag.

Mean colors and magnitudes for the Pal~3 RR Lyrae variables were  
obtained by constructing their light curves. This was achieved by 
matching the Borissova et al. (1998) and Stetson et al. (1999) 
photometries, which allowed light curve types, pulsation periods and 
mean properties to be evaluated (J. Borissova 2000, priv. comm.). While 
these cannot yet be considered the definitive light curves for the 
Pal~3 variables, they certainly represent a substantial improvement in 
comparison with Stetson et al., where only a {\em global} 
``mean" color/magnitude 
was provided for the cluster variables. Importantly, it seems particularly  
clear that: i)~All of the Pal~3 RR Lyrae variables pulsate in the 
fundamental (ab) mode; ii)~All of the Pal~3 RR Lyrae variables are 
redder than $(V-I)_0 \simeq 0.41$~mag: this is the blue limit for RRab 
pulsation in M3, the prototypical Oosterhoff type I cluster (Bakos \& 
Jurcsik 2000; Corwin \& Carney 2001). Note that RRab's are even redder 
in Oosterhoff type II globulars (e.g., Bono, Caputo, \& Marconi 1995).  

From Figs.~6a,b, it is immediately obvious that the HB of Pal~3 is much 
bluer than the HBs of Pal~4 and Eridanus (whose mean combined color is 
indicated, in these plots, by a vertical dotted line, from Catelan 2000). 
However, blue-HB stars appear to be entirely absent from all these 
outer-halo globulars. 
Pal~3 appears to be just slightly more metal-poor than Pal~4 and Eridanus 
(see Stetson et al. 1999 for a detailed discussion). 
The small metallicity difference between Pal~3 and Pal~4/Eridanus suggests 
that metallicity differences alone cannot explain the difference in HB 
color between these clusters as seen in Figs.~6a,b. 
As a matter of fact, Burbidge \& Sandage (1958), in their analysis
of Pal~3 and Pal~4 (which they call the ``10$^{\rm h}$" and the
``11$^{\rm h}$" clusters in their paper, respectively) had already
pointed out that the Pal~3 CMD is ``more normal" than that of
Pal~4 ``because the HB extends to bluer colors." In spite of that
remark, and the existence of several RR Lyrae variables in the cluster
(unlike in Pal~4 and Eridanus) notwithstanding, Pal~3
has usually been considered as good an example of a ``second-parameter"
cluster as any of the other red-HB globulars in the outer halo. Our
qualitative analysis of the CMD presented in Fig.~6 suggests, in line
with Burbidge \& Sandage, that {\em Pal~3 is a much less extreme
second-parameter cluster than Pal~4 and Eridanus}. As we shall see
below, a detailed quantitative analysis confirms that this is indeed
the case.

\begin{figure*}[ht]
 \centerline{\epsfig{file=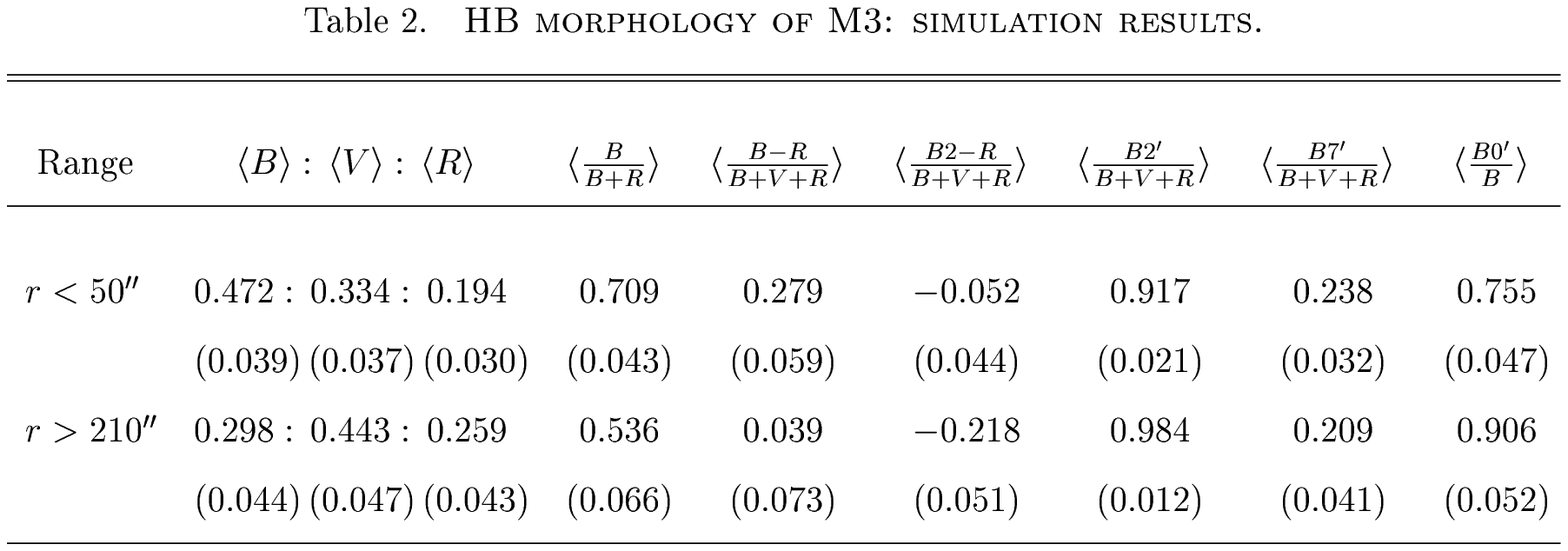,height=1.75in,width=4.75in}}
\end{figure*}

As already pointed out, 
the RR Lyrae population of Pal~3 has not been extensively investigated
to date. However, the currently known RR Lyrae variables in the cluster
all appear to be of the ab-type---i.e., fundamental pulsators (Borissova
et al. 1998; Borissova 2000, priv. comm.; Stetson et al. 1999 and 
references therein). RRab Lyrae variables are well known to be redder 
than RRc variables, implying that the HB of Pal~3, which lacks either 
blue HB or RRc stars, has a narrow dispersion in color. Note that this 
argument independs on the reliability of the individual RR Lyrae 
colors. The mean color of 
the Pal~3 HB stars, variable and non-variable, appears to coincide with
the red edge of the instability strip, at
$(V-I)_0 \approx 0.58$~mag (see Figs.~6a,b). We shall argue, in the next 
section, that the three fairly bright stars at the red extension of the 
HB, at $(V-I)_0 \approx 0.8 - 0.85$~mag, are red HB stars on their way 
to the asymptotic giant branch (AGB), even though the average expected 
number of such evolved stars is $\sim 1$. Taking this into consideration 
in the number counts, we conclude that the most representative number
ratios and mean color describing the HB morphology of Pal~3 are
$B:\,V:\,R = 0.00:\,0.50:\,0.50$ and 
$\langle (V-I)_0 \rangle_{\rm HB} = 0.583 \pm 0.028$~mag, 
respectively.  

Gratton \& Ortolani (1984) suggest that
``one or two stars may be present on the blue side of the RR Lyrae strip" 
of Pal~3. Even though these two stars are very likely cluster members (K. 
Cudworth 2000, priv. comm.), independent photometry by K. Cudworth and by 
S. Ortolani (2000, priv. comm.) shows that they are both variable stars 
(at least one of which almost certainly being an RR Lyrae variable), and 
{\em not} blue-HB stars as had originally been suggested. In fact,  
Ortolani has imaged a Pal~3 field of about $5\arcmin \times 5\arcmin$ 
using the ESO-NTT in 1993, not finding {\em any} blue HB star in the 
cluster. Ortolani points out that Pal~3, in his images, seems to have a 
diameter much smaller than his observed field---namely, no more than 
$\simeq 1.5\arcmin$. 
Independently, Borissova (2000, priv. comm.) has reduced $B$ 
and $V$ images taken with the Rozhen telescope and covering a field 
of $5.6\arcmin \times 5.6\arcmin$ centered on the Pal~3 center. The 
limiting magnitude, in her study, is 23~mag in $V$ (compare with 
Fig.~6). Her analysis of 25 images taken in $B$ and 20 images taken 
in $V$ also confirms that Pal~3 contains {\em no} blue HB stars. 
Given that the Pal~3 half-light radius is only 
$r_{\rm h} \simeq 40\arcsec$ (Harris
1996), the conclusion that Pal~3 does entirely lack blue 
HB stars seems inescapable. Hence the absence of blue HB stars in the 
HST CMD of Figs.~6a,b is not an artifact of the relatively small field of 
view of WFPC2 (compared to the ground-based ones).

Of course, it would obviously be desirable to have a larger sample of HB 
stars. Note however that even photometry covering the entire cluster out 
to its tidal radius is not likely to yield a large sample. Pal~3's tidal 
radius is $r_{\rm t} = 4.81\arcmin$ (Harris 1996); a simple integration 
assuming a King profile suggests that the number of HB stars in the 
range not covered by the Borissova (2000, priv. comm.) or 
Ortolani (2000, priv. comm.) photometries mentioned 
in the previous paragraph---i.e., in between about $3\arcmin$ and 
$5\arcmin$---is very small, of order a few percent of the total 
inside $r \sim 3\arcmin$ only, or less than one HB star. On the 
other hand, it is possible to estimate the ``total" number of 
Pal~3 HB stars from Eq.~(14) in Renzini \& Buzzoni (1986), which 
gives the expected number of stars in any given post-main sequence 
stage as a function of the duration of that stage and of the total 
luminosity of the sampled population. Adopting a lifetime of $10^8$~yr 
for the HB, and a total integrated magnitude $M_V = -5.6$~mag for Pal~3 
(from Harris 1996), we find that $N_{\rm HB,\,tot}^{\rm Pal~3} \approx 30$. 
This number seems surprisingly large in comparison with the number of HB 
stars detected in the cluster so far. We have no explanation at present 
for this discrepancy, but believe that it is highly unlikely that the 
answer will be found in the form of blue HB stars outside the currently 
observed cluster regions. 

\vskip 0.25in
\section{Theoretical Framework}
In this section, our goal is to determine the mean HB mass and HB mass 
dispersion parameters
($\langle M_{\rm HB} \rangle$,~$\sigma_M$) in the simulations
that best account for the observed characteristics of the M3 (\S3.1) 
and Pal~3 (\S3.2) CMDs in the HB region. 

The basic theoretical framework adopted (HB and RGB
evolutionary tracks, color transformations) is the same as in Catelan
(2000 and references therein), and need not be repeated here; we refer
the interested reader to that paper for further information. Our
employed HB synthesis code, {\sc sintdelphi}, is the latest version
of the Catelan et al. (1998 and references therein) code, 
which includes several numerical
improvements and has now been translated to {\sc delphi}.

Our most important working hypothesis is that age is {\em the} second
parameter of HB morphology. We also assume, in line with previous
authors, that M3 and Pal~3 represent a ``bona fide" second-parameter 
pair, thus adopting the same metallicity for the two clusters---which 
appears to be supported by the available data (see Stetson et al. 1999 
and references therein). In particular, Armandroff, Da Costa, \& Zinn 
(1992) have measured a metal abundance for Pal~3 using the Ca{\sc ii} 
triplet method, and found ${\rm [Fe/H]} = -1.57\pm 0.19$~dex on the 
Zinn \& West (1984) scale. This value is to be compared with 
${\rm [Fe/H]} = -1.57$ for M3, as provided by Harris (1996), on the 
same scale. It should be noted that the implied identity between the 
Pal~3 and M3 metal content will eventually have to be placed on a 
firmer footing with the use of high-dispersion spectroscopy; in 
particular, a much smaller error bar than provided for Pal~3 in the 
Armandroff et al. study will be needed to verify the assumption that 
Pal~3 and M3 have indeed the same metallicity.  

Again as in Catelan (2000), we assume that the underlying zero-age HB 
mass distributions can be characterized by Gaussian deviates (see 
Catelan et al. 1998 for a detailed discussion).

%
\begin{figure*}[t]
 \figurenum{6}
 \caption{Panel a) shows the 
          de-reddened HST-WFPC2 CMD for Palomar~3. 
          As indicated, a reddening of $E(V-I) = 0.055$~mag
          has been assumed (based on Schlegel et al. 1998).
          RR Lyrae variables are indicated by circled dots.
          The vertical dotted line indicates the mean
          color of the bulk of the HB populations in the
          outer-halo GCs Pal~4 and Eridanus,
          $\langle (V-I)_0 \rangle = 0.78$~mag (see Catelan 2000).
          Panel b) is identical to panel a), except that in this 
          case a zero-age HB for $Z = 0.001$ (dashed line) is 
          overplotted on the data, as is an evolutionary track for 
          an HB star with the same metallicity and 
          $M = 0.656\,M_{\odot}$, after a suitable shift in $V$ by 
          19.67~mag. 
          In panels c) through h), a random sample of CMD simulations 
          for this cluster (``best-fitting case") is provided for 
          comparison. Note that the simulations do not include the 
          AGB phase. 
         }
\end{figure*}

%
\begin{figure*}[ht]
 \figurenum{7}
 \centerline{\psfig{figure=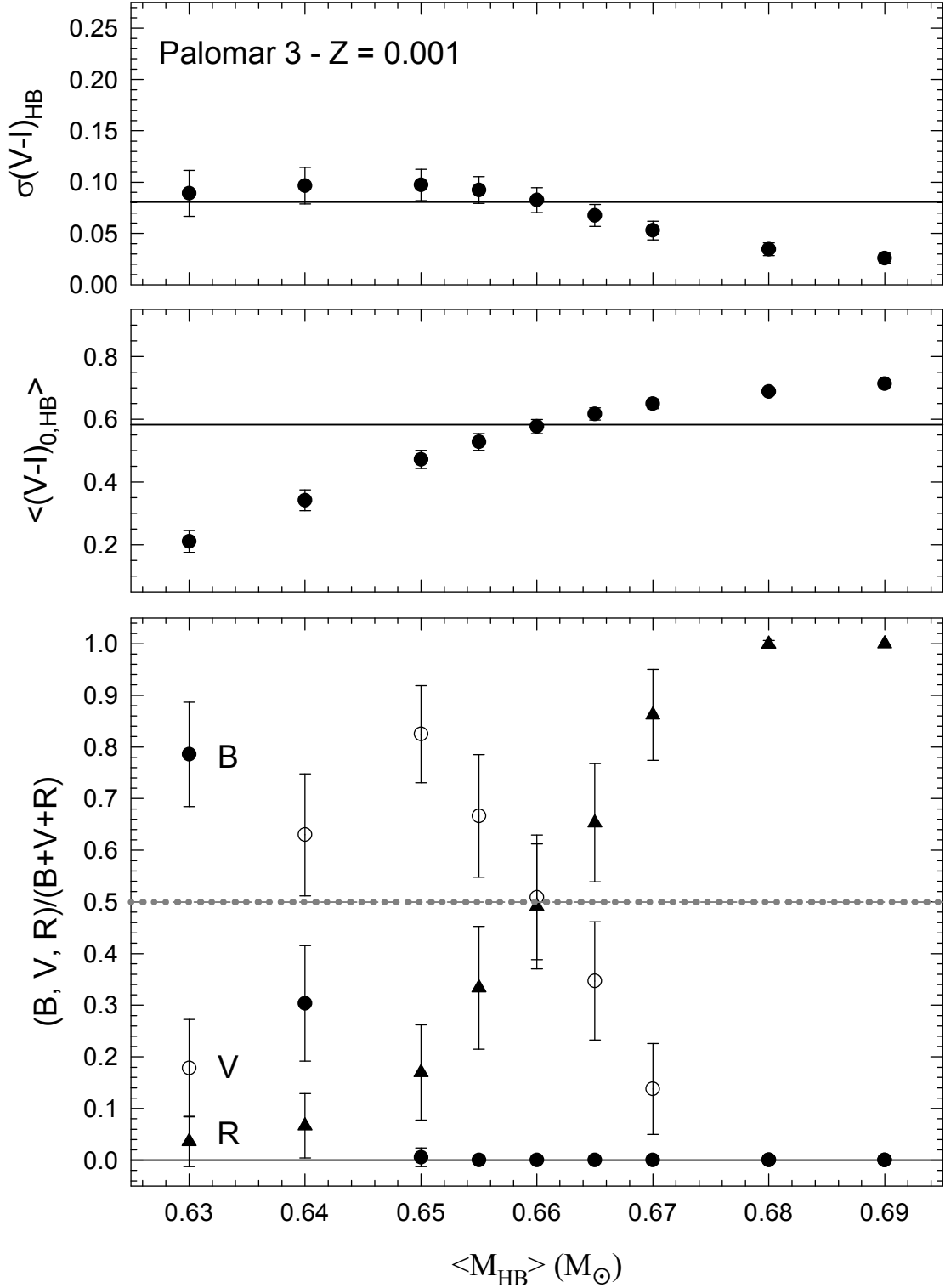}}
 \caption{Illustration of the dependence of several HB morphology 
          parameters on $\langle M_{\rm HB}\rangle$ for Pal~3. 
          The mass dispersion parameter was held fixed at a value 
          $\sigma_M = 0.0035 \, M_{\odot}$, as found appropriate for   
          the cluster.  
          Each datapoint was obtained from an average of 1000 HB 
          simulations. Horizontal lines indicate the observed 
          values. In the bottom panel, the solid line 
          indicates $B/(B+V+R)$, whereas dotted and dashed gray 
          lines (which actually overlap) 
          denote $V/(B+V+R)$ and $R/(B+V+R)$, respectively. 
          In the top panel, the standard deviation in color of 
          the HB distribution, $\sigma(V-I)_{\rm HB}$, is given; 
          the middle panel shows the mean HB color, 
          $\langle (V-I)_{\rm 0,HB} \rangle$. 
         }
\end{figure*}

%
\begin{figure*}[ht]
 \figurenum{8}
 \centerline{\psfig{figure=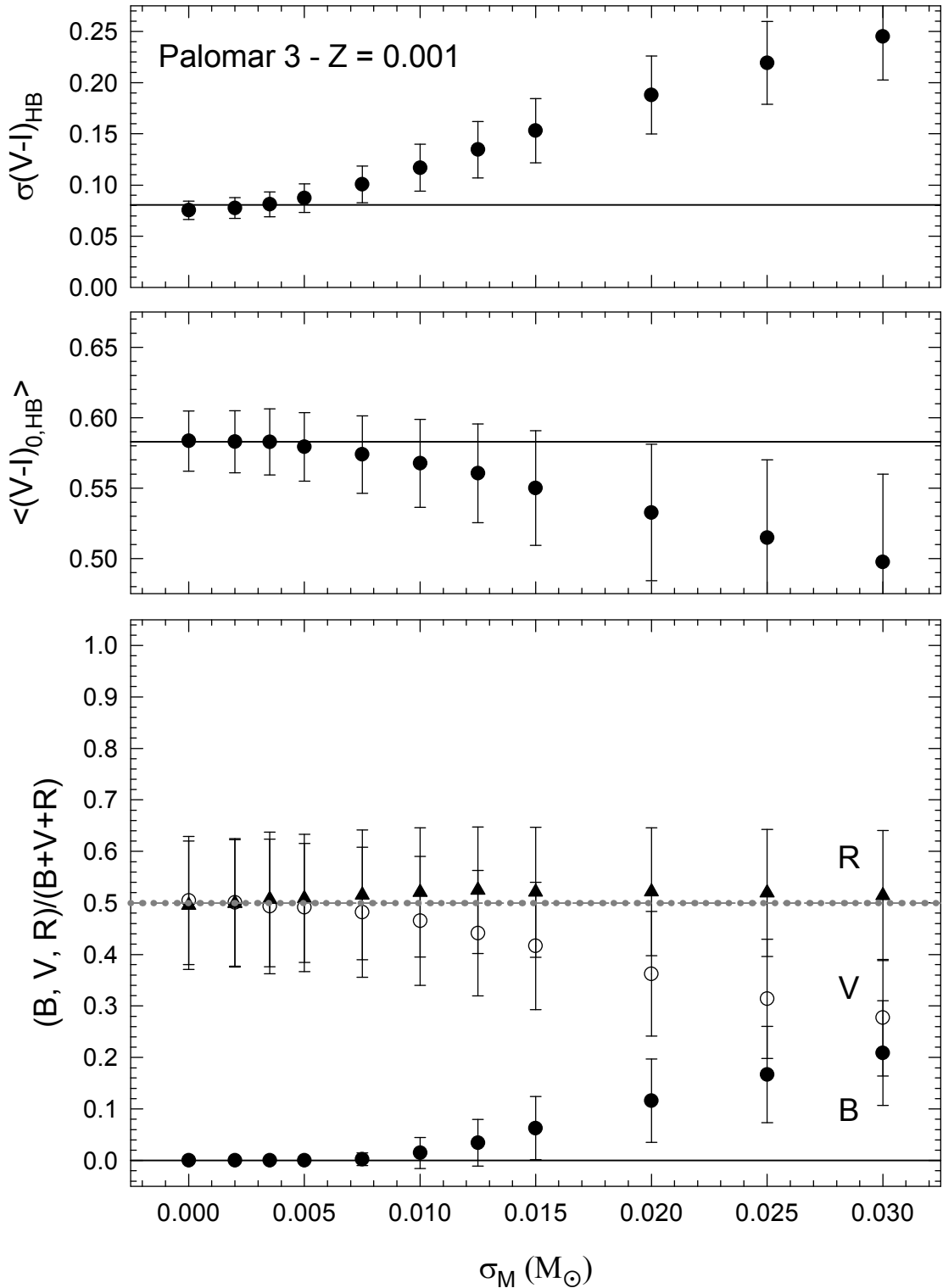}}
 \caption{As in Fig.~7, but for $\sigma_M$. In these computations,  
          the mean HB mass was held fixed at a value 
          $\langle M_{\rm HB}\rangle = 0.661 \, M_{\odot}$, as 
          found appropriate for Pal~3. 
         }
\end{figure*}

%
\begin{figure*}[t]
 \figurenum{9}
 \caption{Same as in Fig.~6, except that in panels c) through h) 
         a random sample of CMD simulations for Pal~3 is provided 
         in the {\em wide, M3-like} ($\sigma_M = 0.02\,M_{\odot}$) 
         mass dispersion case. Note that, 
         unlike in the observed CMD (panels a and b), all of 
         the simulations displayed here
         contain one or more stars bluer than $(V-I)_0 = 0.40$~mag, 
         including at least one bona-fide blue-HB star.
         }
\end{figure*}

%
\begin{figure*}[t]
 \figurenum{10}
 \caption{
         Same as in Fig.~9, except that in this case a subset of the
         wide, M3-like ($\sigma_M = 0.02\,M_{\odot}$) 
         mass dispersion simulations with $B=0$ was
         randomly selected and is displayed in panels c) through h).
         Note that, 
         unlike in the observed CMD (panels a and b) and in the best-fitting 
         simulations (Figs.~6c--h), all of the simulations displayed here
         contain one or more stars bluer than $(V-I)_0 = 0.40$~mag.
         }
\end{figure*}

\subsection{The Case of M3}
Extensive grids of synthetic HBs have been computed aiming at 
estimating the optimum parameters $\langle M_{\rm HB} \rangle$ (mean 
mass) and $\sigma_M$ (mass dispersion) required to reproduce the
observed HB morphology of M3 at both its bluest and reddest
extremes (see \S2.1).

Random ``observational scatter" has also been included, according to  
the following exponential ``laws" that were fitted to the observational 
data: 

\vskip 0.2in 
\begin{displaymath}
\sigma_B = 0.004 \, {\rm exp}(M_B - 3.7) + 0.025, 
\end{displaymath}

\begin{displaymath}
\sigma_V = 0.004 \, {\rm exp}(M_V - 3.8) + 0.018, 
\end{displaymath}

\begin{displaymath}
\sigma_I = 0.004 \, {\rm exp}(M_I - 3.8) + 0.018. 
\end{displaymath}
\vskip 0.2in

\noindent We find that these formulae reproduce the errors in the 
photometry over the whole range of magnitudes 
more accurately than do similar ones adopting 
Robertson (1974) profiles. 

The width of the theoretical instability strip was assumed to be
$\Delta\,\log\,T_{\rm eff} = 0.085$, which leads to a very good
agreement with the empirical blue and red edges of the M3 RR Lyrae
strip as reported by Sandage (1990), Bakos \& Jurcsik (2000), and
Corwin \& Carney (2001). Moreover, the adopted width also leads to
a nice agreement between the model predictions and the observed red
edge of the instability strip in the case of Pal~3, which is 
particularly important for our present purposes (see below).

%
\begin{figure*}[t]
 \figurenum{11}
 \centerline{\psfig{figure=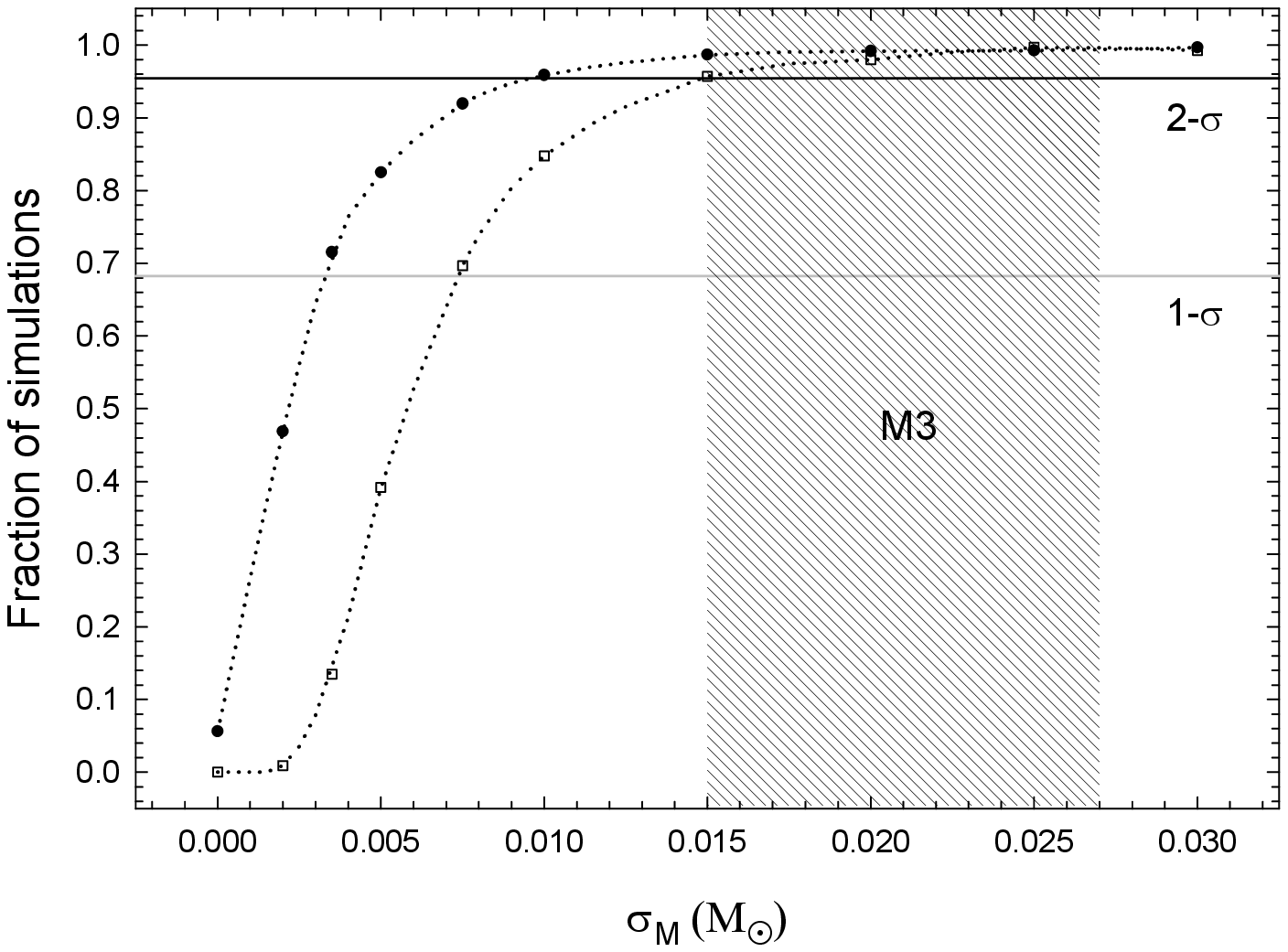}}
 \caption{Fraction of synthetic ``Palomar~3" HB simulations containing  
    at least one star bluer than $(V-I)_0 = 0.4$~mag (open squares)  
    or $(V-I)_0 = 0.45$~mag (filled circles). In these simulations, the 
    mean mass was held fixed at the best-fitting value for Pal~3, 
    $\langle M_{\rm HB} \rangle = 0.661\, M_{\sun}$, but the mass 
    dispersion $\sigma_M$ was allowed to vary over a wide range. The  
    acceptable M3 mass dispersion range (see \S3.1) is schematically 
    shown; $\sigma_M$ could actually be larger than indicated for M3 
    if any of the rejected ``ER" or (especially) ``EB" stars (\S2.1)  
    turn out to be ``bona fide" HB stars. This figure provides an argument, 
    based solely on its lack of sufficiently blue stars, that the Pal~3  
    mass dispersion on the HB is smaller than M3's at the $>2$--$\sigma$  
    level. Note that each data point corresponds to a series of $10,\!000$ 
    Monte Carlo simulations. 
  }
\end{figure*}

We have computed series of 1000 synthetic HBs assuming an
overall number of HB stars $B+V+R = 174$ and 103 for the
``blue" and ``red" HB cases, respectively. These numbers
correspond to the actual number counts in the
inner and outer regions of the cluster, respectively
(see Table~1), and were allowed to fluctuate following the
Poisson distribution. 

The number ratios $B:\,V:\,R$ provided in
Table~1, as well as the dispersion in color in the observed HBs,
allowed us to iteratively fine-tune the model simulation parameters
($\langle M_{\rm HB} \rangle$,~$\sigma_M$)
until the best combinations were obtained. Figures~2 and 3 illustrate 
the dependence of the employed HB morphology parameters on the 
mean HB mass and HB mass dispersion, respectively, for the inner 
regions of the cluster. Figures~4 and 5 are analogous and refer
to the outer regions of M3. Such figures 
were also used to obtain approximate estimates of the corresponding 
errors in the best-fitting values. We found that the following 
provides the best match to the observed HB morphology of M3 in its 
inner regions:

\vskip 0.2in 
\begin{displaymath}
      \langle M_{\rm HB} \rangle = 0.6370\pm 0.0025\, M_{\sun},\,\,\,
                       \sigma_M = 0.023\pm 0.004\, M_{\sun};
\end{displaymath}
\vskip 0.2in 

\noindent whereas, for the outer regions, the best match is found for:

\vskip 0.2in 
\begin{displaymath}
      \langle M_{\rm HB} \rangle = 0.6457\pm 0.0025\, M_{\sun},\,\,\,
                       \sigma_M = 0.018\pm 0.003\, M_{\sun}.
\end{displaymath}
\vskip 0.2in

\noindent Such combinations lead, in the mean, to the HB morphology
parameters displayed in Table~2 (where the numbers in parentheses
represent the standard deviation of the mean over the corresponding
set of 1000 simulations). Note the nice agreement between the observed
(Table~1) and theoretical (Table~2) number counts. We also emphasize 
the agreement between observed and simulated values of the Mironov 
(1972) parameter---which, unlike the Lee-Zinn parameter, is immune 
to possible extant uncertainties in the RR Lyrae number counts in 
the innermost regions of M3 observed only with HST. Note also that 
the best-fitting mass dispersion appears to be slightly larger in the 
inner than in the outer regions of M3, although it will immediately 
be recognized that this result does not bear great statistical 
significance. 

Four simulations were randomly drawn from the two sets of
1000 simulations whose mean HB morphology parameters best reproduce
the observed M3 ones over the two radial regions considered.
The corresponding synthetic CMDs are plotted in Figs.~1c,e
[$M_V$, $(V-I)_0$ plane] and Figs.~1d,f [$M_V$, $(\bv)_0$ plane]
for the inner and outer regions, respectively. For each simulation,
the total number of HB stars, as well as the relative fractions
$B:\,V:\,R$, are indicated. The simulations do not include the
AGB phase.

Our simulations allow us to address the question whether the
difference between the ``inner cluster" and the ``outer cluster"
HB populations is significant or not. We have proceeded as follows.
For the set of simulations which best reproduces, in the mean, the
observed number counts in the {\em inner} regions of M3, we have
selected those for which $V > B$ (i.e., more RR Lyrae variables
than blue-HB stars)---a characteristic of the {\em outer}
regions of M3. We find that only about $3\%$ of the ``inner cluster"
synthetic HBs have ``outer cluster" characteristics, with more RR
Lyrae variables than blue-HB stars. Likewise, about $6\%$ of the ``outer
cluster" simulations have ``inner cluster" characteristics, in the
sense that they show $B > V$. Moreover, since the number of RR Lyrae 
and (especially) red HB stars are essentially the same for the inner 
and outer samples (cf. Table~1 and Figs.~1a,b), the difference arises 
mainly from the number of blue HB stars. If the two populations are 
indeed the same our best estimate of the number of blue HB stars in 
each sample is 55.5. If only Poisson statistics were acting the 
observed inner sample is a 3.5--$\sigma$ deviation upward occuring 
simultaneously with a 3.5--$\sigma$ deviation downward in the outer 
sample. Other scenarios are similarly improbable on a purely 
statistical basis. Hence the difference in HB morphology between 
the inner and outermost regions of M3 is clearly of very high 
statistical significance.

\subsection{The Case of Pal~3}

The HB of Pal~3 that we set out to simulate theoretically, from the HST
observations of Stetson et al. (1999), completely lacks blue HB stars
(see Figs.~6a,b). Inspection of Figs.~6a,b clearly shows that the bulk 
of the
HB stars cluster around the red edge of the instability strip, which 
is located at a color $(V-I)_0 \simeq 0.58$~mag. With the
exception of the three bright, likely evolved stars at
$(V-I)_0 \simeq 0.8 - 0.85$~mag, the dispersion in color among the HB
stars appears to be surprisingly small, especially given the strong 
sensitivity of temperature to mass variations at these intermediate 
colors (e.g., Sweigart, Renzini, \& Tornamb\`e 1987). 
This is confirmed by the fact
that all known or suspected RR Lyrae variables in the cluster are
fundamental (i.e., ab-type) pulsators (Borissova et al. 1998; Stetson
et al. 1999; Borissova 2000, priv. comm.).

The adopted procedure in the synthetic computations is entirely
analogous to the one already described in the case of M3. 
``Observational scatter" has been randomly 
included following exponential ``laws" that were fitted to the HST 
data. The best-fitting error formulae are as follows: 

\vskip 0.2in 
\begin{displaymath}
\sigma_V = 0.0008 \,\, {\rm exp}(M_V - 2.5) + 0.0085, 
\end{displaymath}

\begin{displaymath}
\sigma_I = 0.0007 \,\, {\rm exp}(M_I - 1.5) + 0.0075. 
\end{displaymath}
\vskip 0.2in

In these simulations, the
number of HB stars was allowed to fluctuate around the observed number, 
$B+V+R = 16$, following the Poisson distribution.

As a result, we find that the following parameters lead, in the mean,
to the best overall match between the observed and simulated Pal~3 HBs:

\vskip 0.2in 
\begin{displaymath}
      \langle M_{\rm HB} \rangle = 0.661\pm 0.004\, M_{\sun},\,\,\,
                       \sigma_M = 0.0035^{+0.0045}_{-0.0035}\, M_{\sun}.
\end{displaymath}
\vskip 0.2in 

\noindent The mean number ratios and $(V-I)_0$ color for this set of 
simulations is given in Table~3. A sample of six simulations, drawn 
randomly from the pool of 1000 models, is displayed in Figs.~6c--h. 
Comparison between Figs.~6a,b and Figs.~6c--h 
confirms the good agreement between the 
simulations and the observed CMD. Note, in particular, that the mean 
computed color is in excellent agreement with the one based on the 
HST data (\S2.2). The only aspect of the observed CMD which 
may perhaps be argued not to be well matched by the simulations 
is the slope of the HB. However, much of the sloping 
nature of the Pal~3 HB is due to {\em a single} red HB star, at around 
$(V-I)_0 \approx 0.65$~mag, that appears to be significantly fainter 
than the remaining red HB stars; this is certainly not expected on 
the basis of our simulations (see also VandenBerg 2000). The 
comparison between the models and the observations carried out in 
Fig.~6b suggests that, with the exception of this intriguingly faint 
star, there is no obvious discrepancy between the evolutionary 
models and the observations.

\vskip 0.25in
\parbox{0in}{\epsfxsize=3.25in \epsfysize=1.35in \epsfbox{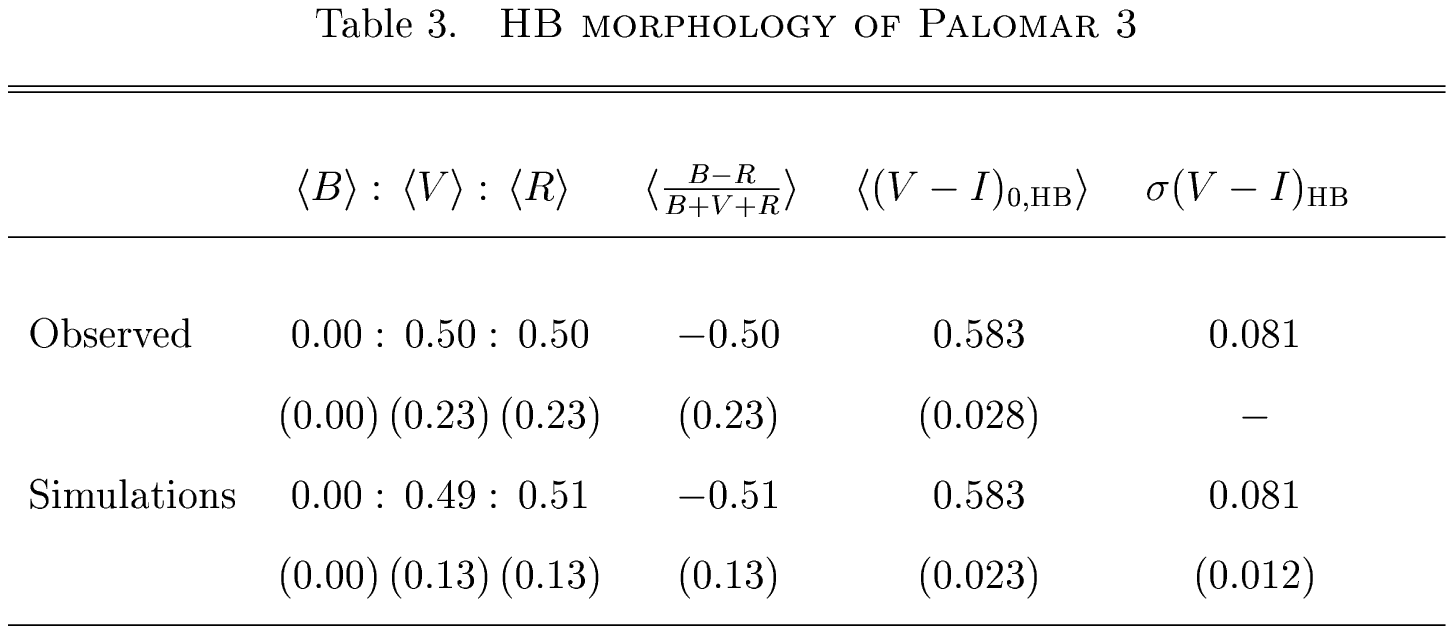}}
\vskip 0.25in

Figures~7 and 8 show the run of several HB morphology parameters for 
Pal~3 as a function of the adopted $\langle M_{\rm HB} \rangle$ and 
$\sigma_M$, respectively. It will be noted that, whereas the HB number 
counts do play an important role when choosing the best-fitting 
$\langle M_{\rm HB} \rangle$ value (Fig.~7, bottom panel), the mean HB 
color, $\langle (V-I)_{\rm 0,HB}\rangle$, is (not unexpectedly) another 
extremely important  quantity constraining the 
$\langle M_{\rm HB} \rangle$ solution. On the other hand, Fig.~8 shows 
that the traditionally employed number counts $B$, $V$, $R$ do not 
pose useful constraints on $\sigma_M$, and neither does the mean HB 
color; in order to estimate the best-fitting $\sigma_M$ value, the 
standard deviation in the HB color, $\sigma (V-I)_{\rm HB}$, is 
clearly the most important quantity (Fig.~8, upper panel). It is 
important to note, at this point, that in order to maximize the  
sensitivity of $\sigma (V-I)_{\rm HB}$ to the HB mass dispersion, 
evolved stars were avoided both in the models and in the observations. 
In the latter case, the three evolved Pal~3 HB stars were not included 
when computing $\sigma (V-I)_{\rm HB}$; in the former, we have decided 
to simply exclude from the computations any star approaching helium 
exhaustion in the core, or (equivalently) the final $10\%$ of HB 
evolution. For these reasons, it is not surprising that 
$\sigma (V-I)_{\rm HB}$ provides a very good diagnostic of mass 
dispersion on the zero-age HB. 

The displayed models
also shed light on the nature of the three red and bright stars lying
to the red of the ``main" Pal~3 HB distribution. For instance, one of 
the displayed simulations, labeled \#942 (Fig.~6c), shows two bright 
and red HB stars, bearing great resemblance to the observed 
distribution. These stars are approaching their final stages of HB 
evolution, on their way up to the AGB. As shown in the figure, the
existence of such bright and red HB stars is not unexpected from our
models (see also Catelan 2000), although we find that the typical
number to be expected in a sample the size of Pal~3's (16 HB stars)
is $\sim 1$. This is due to the relatively short duration of the
final stages of HB evolution close to core helium exhaustion. Though 
suggestive, the three observed bright and red HB stars in Pal~3 cannot 
be considered a serious discrepancy between theory and observations in 
their own right. 

One particularly interesting result of our model computations is the
extremely small value of $\sigma_M$ that we have obtained for Pal~3.
Such a small mass dispersion is largely a consequence of: i)~The
lack of blue HB stars in the cluster; ii)~The lack of first-overtone
RR Lyrae variables (RRc's). One may, however, question the statistical
significance of this result, given that the Pal~3 field that we are
modelling has a mere 16 HB stars. For this reason, we have computed
a new set of simulations for Pal~3 with the same value of
$\langle M_{\rm HB} \rangle = 0.661\, M_{\sun}$ as derived above,
but changing the mass dispersion to a value that appears more
appropriate for M3, namely: $\sigma_M = 0.02\, M_{\sun}$ (see \S3.1).

A random sample of six simulations drawn from the so computed pool
of 1000 models is displayed in Fig.~9. It is readily apparent that
the typical Pal~3 simulation with an M3-like mass dispersion on the
HB shows an HB distribution which does {\em not} provide a good match
to the Pal~3 CMD. In particular, we find that {\em only $12.3\,\%$ of 
these simulations entirely lack blue-HB stars}, as in the HST CMD. Even 
in these cases, however, good morphological agreement with the observed
Pal~3 CMD is generally not obtained. To illustrate this, a random sample 
of six simulations drawn from the (already small) subset of simulations
with $B=0$ is shown in Figs.~10c--h. None of the examples shown has a 
narrow HB color dispersion resembling the observed one (compare 
with Figs.~10a,b), and RRc stars are often present in these simulations 
(as indicated by the instability strip being populated to bluer
colors than in Figs.~6c--h, where very few RRc's are present). As a 
matter of fact, the bluest of the stars plotted in Figs.~10a,b has a 
color $(V-I)_0 = 0.46$~mag, whereas all of the simulations displayed 
in Figs.~10c--h contain stars bluer than $(V-I)_0 = 0.40$~mag. From the 
entire pool of simulations for this M3-like mass dispersion case, we 
indeed find that as many as $98.4\%$ contain one or more stars bluer 
than $(V-I)_0 = 0.40$~mag; this percentage increases to $99.1\%$ if 
one considers $(V-I)_0 = 0.45$~mag as the ``cutoff color" on the 
blue side. We conclude that the small HB mass dispersion that we 
found for Pal~3 is most likely an intrinsic characteristic of Pal~3, 
and not a chance result due to statistical fluctuations. This conclusion  
is also strongly supported by the upper panel of Fig.~8, which clearly 
shows not only that the HB mass dispersion of Pal~3 is smaller than 
$0.01\,M_{\odot}$---but also that it is, perhaps somewhat surprisingly, 
{\em entirely consistent with zero}. Such a small mass dispersion 
for the Pal~3 HB is due to the fact that the HB 
evolutionary tracks, at the Pal~3 mean color, already encompass a 
significant range in color at this metallicity. This is clearly 
demonstrated by Fig.~6b, which shows 
an evolutionary track from Catelan et al. (1998) for an HB star with a 
mass very similar to Pal~3's best-fitting $\langle M_{\rm HB}\rangle$ 
overplotted on the observed CMD. 

To put these results on a firmer statistical basis, we have computed 
many additional sets of simulations in which the mean HB mass was held
fixed at the ``optimum" value for Pal~3, namely 
$\langle M_{\rm HB} \rangle = 0.661\, M_{\sun}$, but allowing the 
mass dispersion $\sigma_M$ to vary between $0.0\,M_{\odot}$ (``single 
HB evolutionary track case") and $0.035\,M_{\odot}$. For each $\sigma_M$
value, $10,\!000$ Monte Carlo simulations with 16 HB stars 
in each (as observed), plus Poisson noise, were computed, 
and the number of those simulations showing one or more HB star bluer 
than the ``cutoff colors" $(V-I)_0 = 0.40$~mag and $(V-I)_0 = 0.45$~mag
obtained. The results of these simulations are summarized in Fig.~11, 
where 
the percentage of simulations containing one or more star bluer than 
the indicated cutoffs is plotted as a function of the mass dispersion 
$\sigma_M$. The range of mass dispersion values found for M3 (from 
\S3.1) is also 
schematically indicated; note that these may be underestimates, in 
case any of the rejected ``ER" and ``EB" stars (\S2.1), particularly 
the latter, turn out to be ``bona fide" HB stars. 

Figure~11 clearly confirms that the lack of bluer HB stars in the Pal~3 
HST-WFPC2 CMD implies that the HB mass dispersion in this cluster is 
smaller than M3's at the $>2$--$\sigma$ level. This is true even when 
the cutoff color is placed at $(V-I)_0 = 0.40$~mag, which is 0.06~mag 
bluer than the reddest point on the HST-WFPC2 CMD. The choice of such 
a significantly bluer cutoff color was motivated by the fact that, in 
spite of recent efforts,  the Pal~3 RR Lyrae light curves are still 
not very well determined, 
so that their individual colors remain somewhat uncertain. However, 
we again emphasize that 
all RR Lyrae stars found to date in Pal~3 appear to be fundamental 
pulsators. From Fig.~2 in Bakos \& Jurcsik (2000) and from Corwin \& 
Carney (2001) it is clear, in 
turn, that all RRab stars in M3 (a prototypical Oosterhoff type I 
cluster) are redder than $(V-I)_0 \simeq 0.41$~mag; RRab's are even 
redder in Oosterhoff type II globulars (e.g., Bono et al. 1995).

In conclusion, it appears clear that the mass dispersion on the Pal~3 
HB is very small, being not only intrinsically smaller than M3's 
but also indistinguishable from zero.

\begin{figure*}[ht]
 \centerline{\epsfig{file=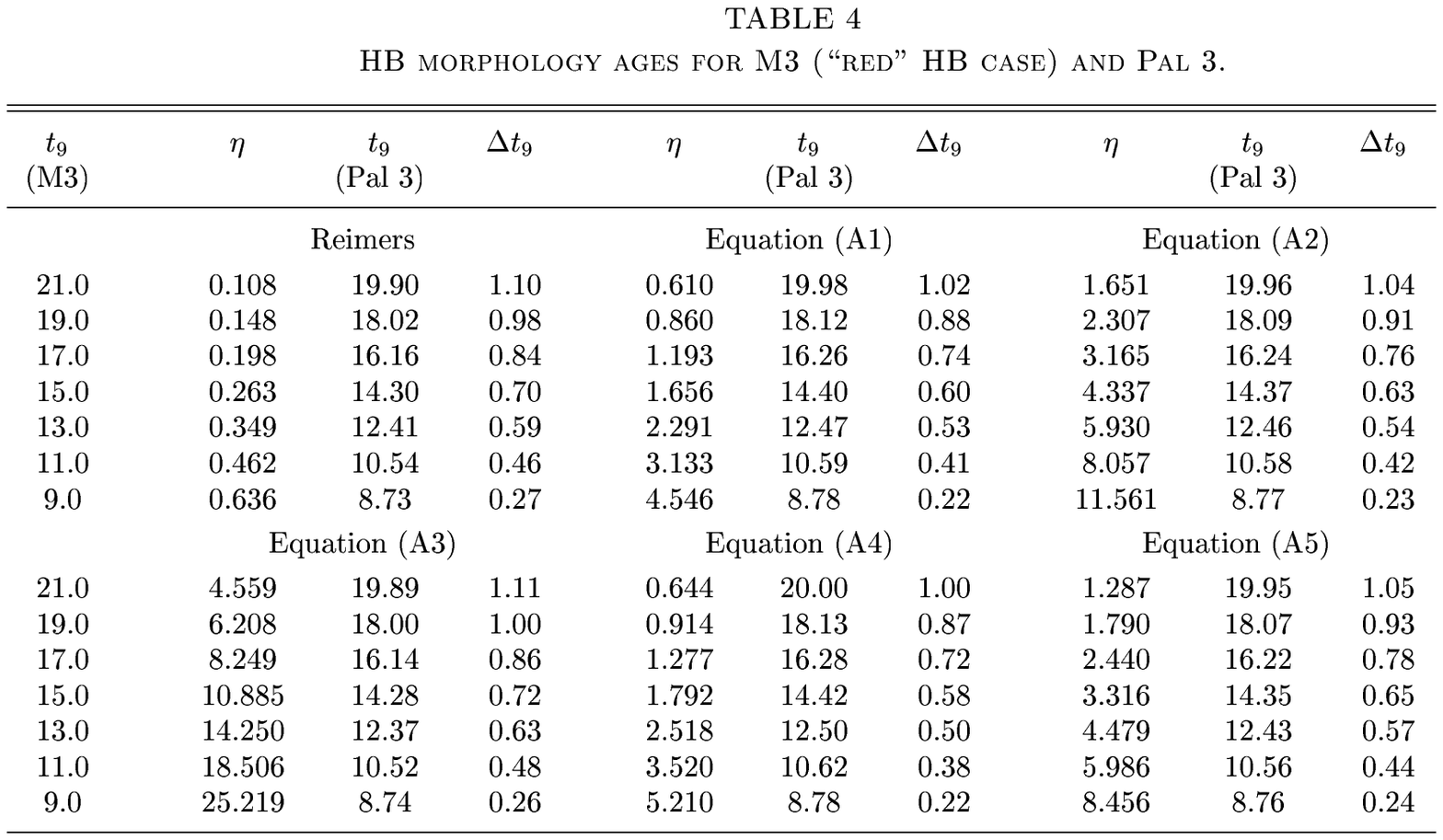,height=3.25in,width=5.35in}}
\end{figure*}

\begin{figure*}[ht]
 \centerline{\epsfig{file=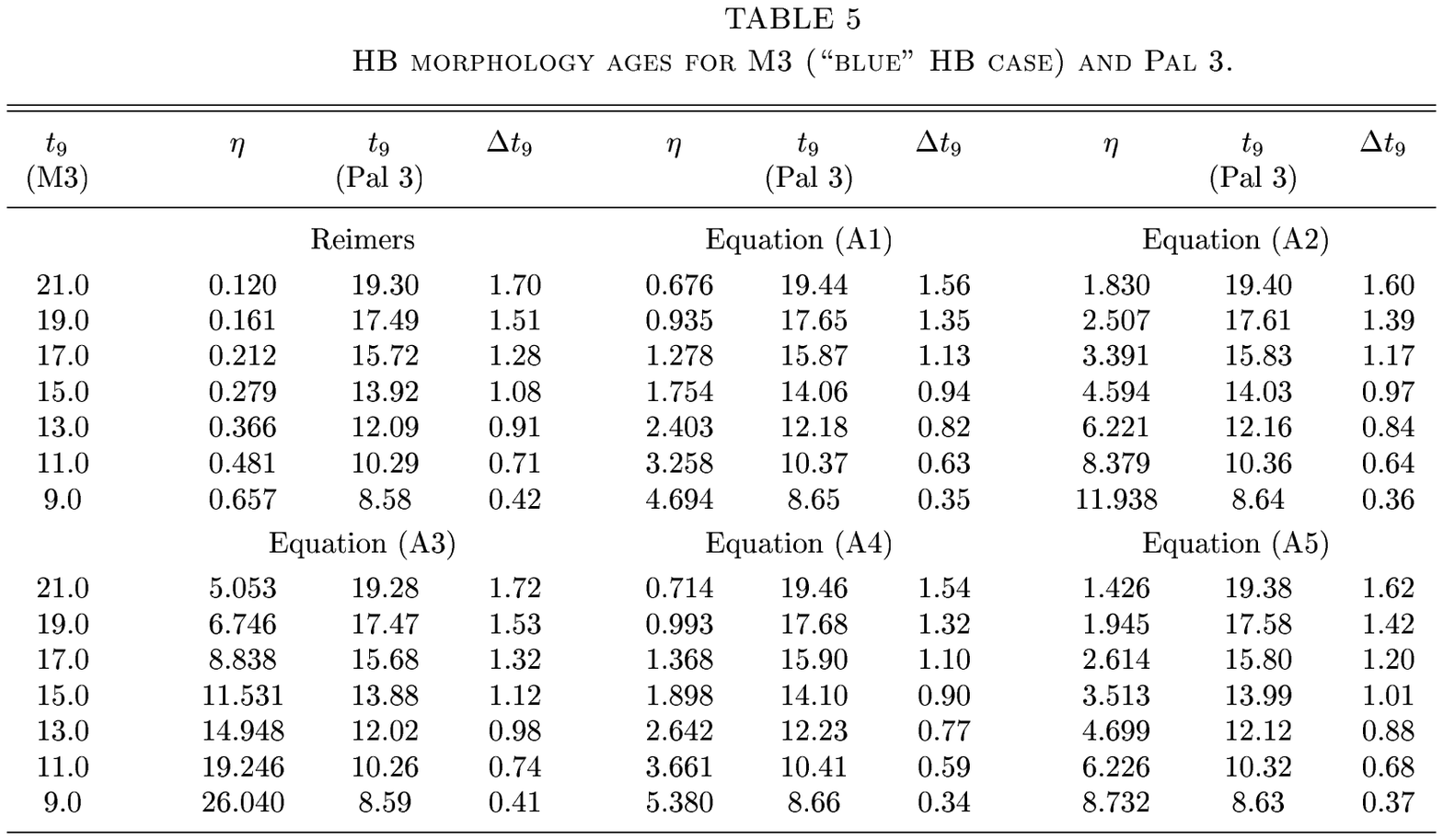,height=3.25in,width=5.35in}}
\end{figure*}

%
\begin{figure*}[t]
 \figurenum{12}
 \centerline{\psfig{figure=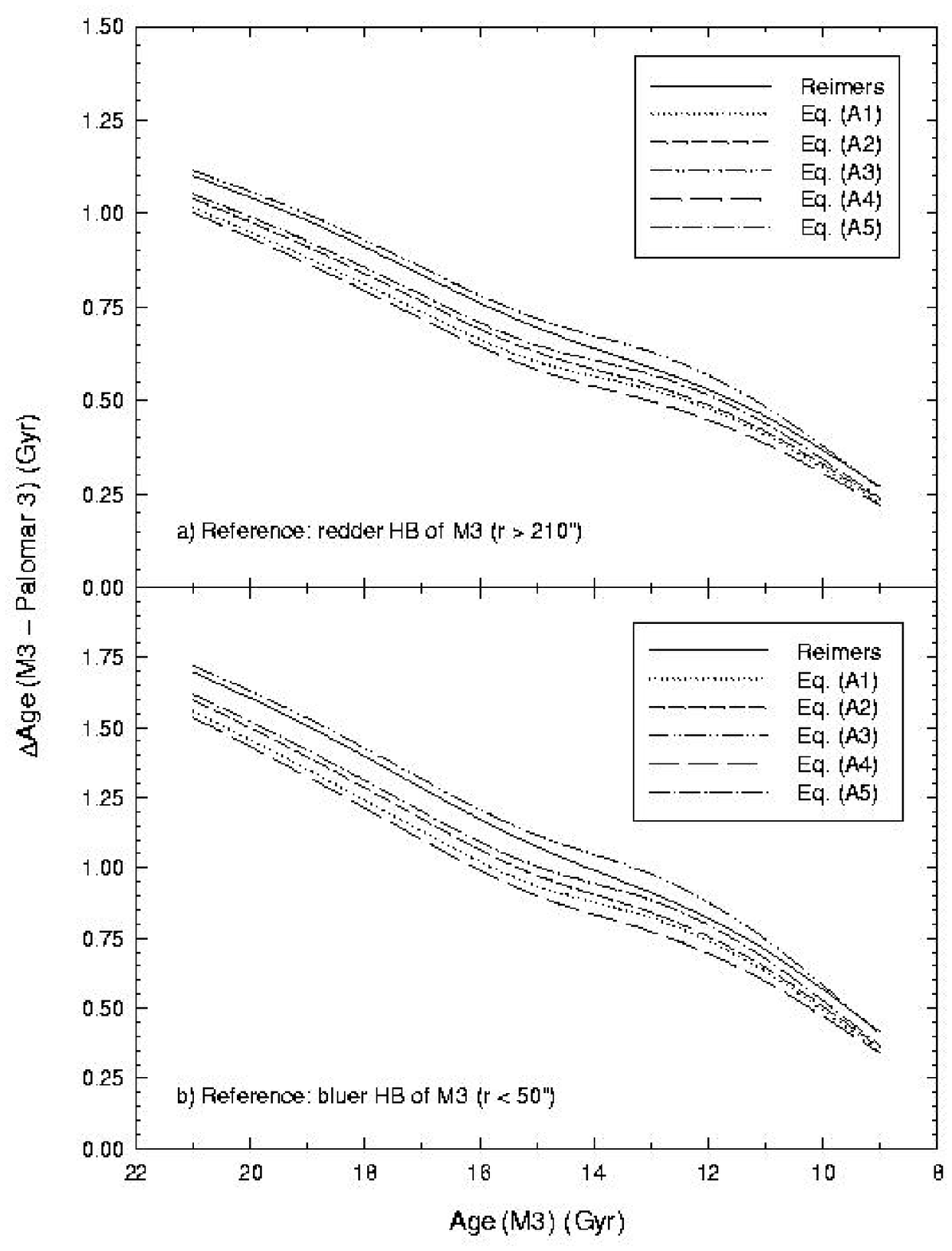}}
 \caption{The difference in age
    $\Delta{\rm Age}({\rm M3-Pal~3})$ (in Gyr),
    derived for the several indicated mass loss formulae
    (see the Appendix in Catelan 2000),  
    is plotted as a function of the M3 age (also in Gyr).
    In the lower panel, the best-fitting M3 simulations
    for the innermost ($r < 50\arcsec$) radial range are
    considered (bluest HB type), whereas in the upper panel
    the best-fitting M3 simulations for the outermost
    ($r < 210\arcsec$) radial range are employed (reddest
    HB type). 
  }
\end{figure*}

\vskip 0.25in

\section{Estimating Relative Ages from Differences in the HB Morphology}

In order to estimate the relative ages required to produce the
relative HB types of M3 and Pal~3, we follow the same approach
described in Catelan (2000) in the case of M5 and Pal~4/Eridanus.
As in that paper, we shall evaluate the effects of an age-dependent
mass loss on the RGB, as implied by the several different analytical
formulae discussed in Catelan's Appendix.
As argued by Catelan, at the present time it does not appear
possible to strongly argue in favor of any of his recalibrated
equations over the others, so that a safer approach is to use
all of them simultaneously whenever an estimate of the amount
of mass loss on the RGB is needed. We briefly recall here that
equations~(A1) through (A4) of Catelan are ``generalized,"
empirically recalibrated variations of the
analytical mass loss formulae previously suggested by Reimers
(1975a, 1975b), Mullan (1978), Goldberg (1979), and Judge \&
Stencel (1991), respectively.
In this paper, RGB mass loss was estimated on the basis of the
RGB models of VandenBerg et al. (2000) for a chemical composition
${\rm [Fe/H]} = -1.54$, $[\alpha/{\rm Fe}] = +0.3$.

We follow the same approach described in Catelan (2000) to measure
the age difference between the two clusters, as required from
their HB morphologies under the assumption that age is the second
parameter. Given the variation in HB morphology as a function
of radial position in the case of M3, we are forced to study the
M3 inner regions (bluer HB) and outer regions (redder HB)
as limiting cases for deriving the HB morphology-based age
difference between M3 and Pal~3; we assume that the most
appropriate age difference estimate will lie somewhere in
between these limits. In particular, the ``integrated" 
HB morphology of M3 can be characterized by the number ratios 
$B:\,V:\,R = 0.39\,: 0.40\,: 0.21$, which would correspond to 
a synthetic HB falling in between the two limiting cases 
considered above (compare with Tables~2 and 3). 

For each of the Catelan (2000) mass loss formulae, and also for
Reimers' (1975a, 1975b) formula, the resulting $\eta$ values and
``HB morphology ages" for M3 and Pal~3 are given in
Table~4 (``red HB" case for M3) and Table~5 (``blue HB" case for
M3). These results are also shown in Fig.~12, where one
can find our theoretical age difference vs. absolute age curves
for each mass loss formula (see labels) both in the ``red HB"
and ``blue HB" limits for M3 (upper panel and lower panel,
respectively).

VandenBerg (2000), on the basis of  
deep WFPC2 photometry for Pal~3, has recently found a turnoff 
(TO) age 
``of $\approx 11.5$~Gyr for Pal~3, which is to be compared 
with $12-12.5$~Gyr for M3" (see his \S2.4.2)---implying a 
{\em favored} age difference for the M3--Pal~3 pair of  
$\approx 0.5-1$~Gyr. Comparing such a TO age difference  
with the HB morphology-based results displayed in Fig.~12, we 
find that the relative TO ages {\em favored} by VandenBerg are 
largely consistent with the relative HB types of the M3 and Pal~3 
HBs. 

On the other hand, Stetson et al. (1999) state that ``the 
[deep WFPC2] data are consistent with an age difference of some 
2~Gyr between M3 and Pal~3 (Pal~3 younger) if their abundances 
are the same." Fig.~12 clearly shows that such a larger TO age 
difference is not favored by the present models under the same 
assumption as adopted by Stetson et al., namely: identical 
abundances. Age differences for this pair in the 2~Gyr range 
might still be reconciled with the 
HB morphologies of the studied clusters, provided other ``second 
parameters" vary in such a way as to ``compensate" the effects of 
such relatively large age differences upon
the predicted HB morphologies. The $\alpha$-element abundance 
is one such plausible candidate; indeed, [$\alpha$/Fe]
has been observed to be lower in the ``young," loose GCs Pal~12
and Ruprecht~106 (Brown, Wallerstein, \& Zucker 1997) than in the
vast majority of Galactic GCs (Carney 1996 and references therein).
However, a lower [$\alpha$/Fe] for Pal~3 would drive a {\em bluer}
HB type for the cluster, thus decreasing even further the (already
small) age difference needed to account for the difference in HB
morphology between M3 and Pal~3. Hence a lower [$\alpha$/Fe] for
Pal~3 would work in the {\em opposite} sense, from the standpoint
of HB models, to what would be needed to produce agreement between
the present results and a TO age difference of about 2~Gyr. 

Of 
course, it may also be that future observations will reveal that 
the [Fe/H] values of Pal~3 and M3 differ, thus removing these  
clusters from the ``bona fide second parameter pair" category 
which they currently seem to occupy. Empirically, one can see, from 
an HB morphology--metallicity diagram such as the one presented in 
Fig.~1 of Catelan \& de Freitas Pacheco (1993), that an increase 
in [Fe/H] by 0.2~dex (i.e., within the Armandroff et al. 1992 
derived uncertainty) could move the Pal~3 
HB to the red, compared to that of M3, by approximately the right 
amount, even if no age difference is present. (We have, of course, 
ignored, for the sake of argument, both the internal second 
parameter in M3 and the intrinsically narrower mass distribution 
of Pal~3.) Conversely, it is 
conceivable that, if Pal~3 turns out to be more metal-poor than 
M3 by 0.2~dex, a larger TO age difference with respect to this 
cluster will be required in order to reproduce its red HB. 
Unfortunately, our imperfect (at best) 
knowledge of the variation of RGB mass loss 
with metallicity prevents us from carrying out a more meaningful 
analysis of the effects of metallicity variations upon the HB 
morphology of Pal~3; again, the importance of improving our 
knowledge of mass loss in first-ascent giants cannot be 
overemphasized.

\section{Conclusions and Discussion}
In the present paper, we have carried out a detailed analysis of the
``second-parameter pair" of globular clusters M3 and Palomar~3, based 
both on observational data and theoretical models. We have shown that
the HB morphology of M3 is significantly bluer in its inner regions
(observed with HST) than in the cluster outskirts (observed from the
ground), and have presented Monte Carlo simulations which support the
view that this difference in HB type is unlikely to be due to 
statistical fluctuations. This result leads to the conclusion that
some ``local" second parameter(s) is acting between the inner and
outer parts of M3.

Our Monte Carlo simulations also suggest that the dispersion in mass 
of the Pal~3 HB is very small (consistent with zero), which is unlikely 
to be due to a statistical fluctuation in spite of the small sample 
size in the cluster CMD. In this regard, the difference in HB mass 
dispersion between M3 and Pal~3 is clearly 
found to be statistically significant. 

Ignoring the internal second parameter in M3 and the narrow mass
distribution in Pal~3, we can do a traditional analysis searching for a
global second parameter acting between the two clusters. For example,
we can search for an age difference based on the mean color or HB mass
of each cluster.  In this case, under the assumption that the two
clusters have the same metallicity, the relative HB types of M3 and
Pal~3 can easily be accounted for by a fairly small difference in age
between them, of order 0.5--1~Gyr---which is in good agreement
with the relative age measurement, based on the clusters' turnoffs,
recently presented by VandenBerg (2000). Hence age could be the 
primary ``global" second parameter for the pair M3--Pal~3.  

Even though there may be a small age difference between M3 and Pal~3
(Stetson et al. 1999; VandenBerg 2000), the dramatically different HB 
morphologies obvious from a cursory examination of the CMDs appears 
to arise from 
another factor---the small mass dispersion of the Pal~3 HB. For some 
reason the star-to-star differences that lead to 
$\sigma_M \sim 0.02 - 0.03\,M_{\odot}$ in most GCs do not seem to occur 
in Pal~3. What possible mass loss-driving stellar parameter could have 
such a behavior?

M3's internal second parameter adds additional complexity to the
problem of identifying a global second parameter acting between 
M3 and any other 
cluster. It is unlikely that age is the internal second parameter
because over the $>10$~Gyr age of the cluster dynamical mixing should
have eliminated any initial radial separation of two age groups. On
the other hand dynamical separation of stars which have undergone
differential mass loss is unlikely to occur on a time scale comparable
to or less than the HB lifetime. Even if it did, the settling of the
more massive red HB stars to the center would produce a result
opposite that observed. The most plausible internal second parameter
is that some environmental factor leads to enhanced mass loss for the
inner stars. One can at least imagine that stellar encounters could
affect rotation rates which could in turn affect mass loss. If this is
the case other clusters with HBs similar to M3, e.g., M5, might show
an internal second parameter; it would be interesting to investigate 
this possibility further using a combination of HST and ground-based
photometry (as done in this paper for M3). Perhaps the density of the 
stellar medium, 
at least in this case and in Pal~3's, will turn out to be the culprit, 
provided it can be shown that RGB mass loss and the density of stars 
in globular clusters are somehow related (with higher density leading 
to an increase in the overall amount of mass loss on the RGB and in 
the dispersion around the mean value). 
At least in the case of M3, another 
intriguing possibility is that the detected radial variation in 
HB morphology is somehow related to the peculiar radial variations 
observed in the blue stragglers populations in the same cluster 
(Ferraro et al. 1997b). Again we find ourselves in the position 
where a detailed analysis of a second parameter pair does not lead 
to a simple answer.

\acknowledgments
The authors are grateful to P. B. Stetson for kindly supplying the 
Pal~3 HST photometry. J. Borissova's aid in the analysis of the 
Pal~3 RR Lyrae variables is gratefully acknowledged. We also thank 
M. Bellazzini, K. Cudworth, D. Dinescu, S. Ortolani, H. A. Smith, 
and D. A. VandenBerg for useful discussions and/or information. 
Useful comments by an anonymous referee are also gratefully 
acknowledged. Support for M.C. was provided by NASA through
Hubble Fellowship grant HF--01105.01--98A awarded by the Space
Telescope Science Institute, which is operated by the Association
of Universities for Research in Astronomy, Inc., for NASA under
contract NAS~5--26555. F.R.F acknowledges the finacial support 
of the Agenzia Spaziale Italiana and the Ministero della
Universit\`a e della Ricerca Scientifica e Tecnologica to the 
project {\em Dynamics and Stellar Evolution}. R.T.R. is partially 
supported by NASA LTSA Grant NAG~5--6403 and STScI Grant~GO--6607.


\begin{references}

\reference{} Armandroff, T. E., Da Costa, G. E., \& Zinn, R. 1992, \aj, 
   104, 164

\reference{} Bailyn, C. D., Iben, I., Jr. 1989, \apjl, 347, L21

\reference{} Bakos, G. A., \& Jurcsik, J. 2000, in ASP Conf. Ser. Vol.
   203, The Impact of Large-Scale Surveys on Pulsating Star Research,
   ed. L. Szabados \& D. W. Kurtz (San Francisco: ASP), 255

\reference{} Bolte, M., \& Hogan, C. J. 1995, \nat, 376, 399

\reference{} Bono, G., Caputo, F., \& Marconi, M. 1995, \aj, 110, 2365

\reference{} Borissova, J., Catelan, M., Spassova, N., \& Sweigart, A.
   V. 1997, \aj, 113, 692

\reference{} Borissova, J., Spassova, N., Catelan, M., \& Ivanov, V. D.
   1998, in ASP Conf. Ser. Vol.
   135, A Half-Century of Stellar Pulsation Interpretations,
   ed. P. A. Bradley \& J. A. Guzik (San Francisco: ASP), 188

\reference{} Buonanno, R. 1993, in ASP Conf. Ser. Vol. 48, The Globular 
  Cluster-Galaxy Connection, ed. G. H. Smith \& J. P. Brodie (San Francisco: 
  ASP), 131

\reference{} Buonanno, R., Corsi, C. E., Buzzoni, A., Cacciari, C.,
Ferraro, F. R., \& Fusi Pecci, F. 1994, \aap, 290, 69

\reference{} Brown, J. A., Wallerstein, G., \& Zucker, D. 1997, \aj, 114,
   180

\reference{} Burbidge, E. M., \& Sandage, A. 1958, \apj, 127, 527

\reference{} Carney, B. W. 1996, \pasp, 108, 900

\reference{} Carretta, E., Cacciari, C., Ferraro, F. R., Fusi Pecci, F.,
   Tessicini, G. 1998, \mnras, 298, 1005


\reference{} Catelan, M. 2000, \apj, 531, 826

\reference{} Catelan, M., Bellazzini, M., Landsman, W. B., Ferraro, F. R., 
  Fusi Pecci, F., \& Galleti, S. 2001, \aj, submitted 

\reference{} Catelan, M., Borissova, J., Sweigart, A. V., \& Spassova, N.
  1998, \apj, 494, 265

\reference{} Catelan, M., \& de Freitas Pacheco, J. A. 1993, \aj, 106, 1858

\reference{} Corwin, T. M., \& Carney, B. W. 2001, preprint

\reference{} Ferraro, F. R., Fusi Pecci, F., Cacciari, C., Corsi, C. E.,
   Buonanno, R., Fahlman, G. G., \& Richer, H. B. 1993, \aj, 106, 2324

\reference{} Ferraro, F. R., Carretta, E., Corsi, C. E., Fusi Pecci, F.,
   Cacciari, C., Buonanno, R., Paltrinieri, B., \& Hamilton, D. 1997a, \aap,
   320, 757


\reference{} Ferraro, F. R., et al. 1997b, \aap, 324, 915

\reference{} Fusi Pecci, F., \& Bellazzini, M. 1997, in The Third Conference
   on Faint Blue Stars, ed. A. G. Davis Philip, J. W. Liebert, \& R. A.
   Saffer (Schenectady: L. Davis Press), 255

\reference{} Fusi Pecci, F., Bellazzini, M., Cacciari, C., \& Ferraro, F.
   R. 1995, \aj, 110, 1664

\reference{} Fusi Pecci, F., Ferraro, F. R., Bellazzini, M., Djorgovski, S.,
   Piotto, G., \& Buonanno, R. 1993, \aj, 105, 1145

\reference{} Goldberg, L. 1979, \qjras, 20, 361

\reference{} Gratton, R. G., \& Ortolani, S. 1984, \aaps, 57, 177

\reference{} Grebel, E. K. 2001, in ASP Conf. Ser., Microlensing 2000:
   A New Era of Microlensing Astrophysics, eds. J. W. Menzies \& P. D.
   Sackett, in press (astro-ph/0008249)


\reference{} Harris, W. E. 1996, \aj, 112, 1487 (June 22$^{\rm nd}$, 1999
   version)

\reference{} Harris, W. E., et al. 1997, \aj, 114, 1030

\reference{} Judge, P. G., \& Stencel, R. E. 1991, \apj, 371, 357 

\reference{} Lee, Y.-W., Demarque, P., \& Zinn, R. 1994, \apj, 423, 248

\reference{} Majewski, S. R. 1994, \apjl, 431, L17

\reference{} Mironov, A. V. 1972, AZh, 49, 134

\reference{} Mironov, A. V., \& Samus, N. N. 1974, Peremennye Zv\"ezdy,
   19, 337

\reference{} Mullan, D. J. 1978, \apj, 226, 151

\reference{} Ortolani, S., \& Gratton, R. G. 1989, \aaps, 79, 155

\reference{} Reimers, D. 1975a, in Problems in Stellar Atmospheres and
   Envelopes, ed. B. Baschek, W. H. Kegel, \& G. Traving (Berlin:
   Springer-Verlag), 229

\reference{} Reimers, D. 1975b, in Probl\`emes D'Hydrodynamique Stellaire,
   M\'emoires de la Societ\'e Royale des Sciences de Li\'ege, 6e serie,
   tome VIII, 369

\reference{} Renzini, A., \& Buzzoni, A. 1986, in Spectral Evolution of
   Galaxies, ed. C. Chiosi \& A. Renzini (Dordrecht: Reidel), 195

\reference{} Richer, H. B., et al. 1996, \apj, 463, 602

\reference{} Robertson, J. W. 1974, \apj, 191, 67

\reference{} Rood, R. T., et al. 1999, \apj, 523, 752

\reference{} Sandage, A. 1990, \apj, 350, 603

\reference{} Sarajedini, A., Chaboyer, B., \& Demarque, P. 1997,
   \pasp, 109, 1321

\reference{} Sarajedini, A., Geisler, D., Harding, P., \& Schommer, R. 
   1998, \apjl, 508, L37

\reference{} Schlegel, D. J., Finkbeiner, D. P., \& Davis, M. 1998,
   \apj, 500, 525

\reference{} Searle, L., \& Zinn, R. 1978, \apj, 225, 357

\reference{} Stetson, P. B., et al. 1999, AJ, 117, 247

\reference{} Stetson, P. B., VandenBerg, D. A., \& Bolte, M. 1996,
      \pasp, 108, 560

\reference{} Sweigart, A. V., Renzini, A., \& Tornamb\`e, A. 1987, \apj, 
   312, 762

\reference{} van den Bergh, S. 1993, \apj, 411, 178

\reference{} van den Bergh, S. 1999, \aapr, 9, 273

\reference{} van den Bergh, S. 2000, \pasp, 112, 932

\reference{} VandenBerg, D. A. 2000, \apjs, 129, 315

\reference{} VandenBerg, D. A., Swenson, F. J., Rogers, F. J., Iglesias,
   C. A., \& Alexander, D. R. 2000, \apj, 532, 430

\reference{} Zinn, R. 1986, in Stellar Populations, ed. C. A. Norman,
   A. Renzini, \& M. Tosi (Cambridge: CUP), 73

\reference{} Zinn, R. 1993, in ASP Conf. Ser. Vol. 48, The Globular 
   Cluster--Galaxy Connection, ed. G. A. Smith \& J. P. Brodie (San 
   Francisco: ASP), 38 

\reference{} Zinn, R., \& West, M. J. 1984, \apjs, 55, 45

\end{references}
\end{document}